\documentclass{pasa}%

\usepackage{graphicx}
\usepackage{color}
\usepackage{siunitx}
\usepackage{cleveref}

\DeclareSIUnit\parsec{pc}
\DeclareSIUnit\erg{erg}
\DeclareSIUnit\jansky{Jy}
\newcommand{\change}[1]{#1}

\title[MWA Phase~II Science]{Science with the Murchison Widefield Array: Phase~I Results and Phase~II Opportunities}

\makeatletter
\newcounter{MWAaffil}
\newcommand{\MWAaffil}[1]{%
  \ifcsname MWAaffil@#1\endcsname
  \else
    \stepcounter{MWAaffil}%
    \expandafter\xdef\csname MWAaffil@#1\endcsname{\theMWAaffil}%
  \fi
  \csname MWAaffil@#1\endcsname}
\makeatother

\def\builders{
B.~Crosse$^{\MWAaffil{CurtinCIRA}}$,
D.~Emrich$^{\MWAaffil{CurtinCIRA}}$,
T.~M.~O.~Franzen$^{\MWAaffil{Curtin},\MWAaffil{ASTRON2}}$,
L.~Horsley$^{\MWAaffil{CurtinCIRA}}$,
D.~Kenney$^{\MWAaffil{Curtin}}$,
M.~F.~Morales$^{\MWAaffil{UW}}$,
D.~Pallot$^{\MWAaffil{UWA}}$,
K.~Steele$^{\MWAaffil{CurtinCIRA}}$,
S.~J.~Tingay$^{\MWAaffil{Curtin}, \MWAaffil{CAASTRO}, \MWAaffil{INAF}}$,
M.~Walker$^{\MWAaffil{CurtinCIRA}}$,
R.~B.~Wayth$^{\MWAaffil{Curtin}}$,
A.~Williams$^{\MWAaffil{CurtinCIRA}}$,
and C.~Wu$^{\MWAaffil{UWA}}$
}
\def\affils{
\affil{$^{\MWAaffil{ASU}}$School of Earth and Space Exploration, Arizona State University, Tempe, AZ 85287, USA}
\affil{$^{\MWAaffil{Curtin}}$International Centre for Radio Astronomy Research, Curtin University, Bentley, WA 6102, Australia}
\affil{$^{\MWAaffil{ASTRO3D}}$ARC Centre of Excellence for All Sky Astrophysics in 3 Dimensions (ASTRO 3D), Perth, WA 6845, Australia}
\affil{$^{\MWAaffil{Brown}}$Department of Physics, Brown University, Providence, RI 02912, USA}
\affil{$^{\MWAaffil{Tata}}$National Centre for Radio Astrophysics, Tata Institute of Fundamental Research, Pune 411007, India}
\affil{$^{\MWAaffil{UWisc}}$Department of Physics, University of Wisconsin--Milwaukee, Milwaukee, WI 53201, USA}
\affil{$^{\MWAaffil{USydney}}$Sydney Institute for Astronomy, School of Physics, The University of Sydney, NSW 2006, Australia}
\affil{$^{\MWAaffil{UCBerkeley}}$University of California, Berkeley, Astronomy Department, 501 Campbell Hall \#3411, Berkeley, CA 94720, USA}
\affil{$^{\MWAaffil{UWA}}$International Centre for Radio Astronomy Research, University of Western Australia, Crawley 6009, Australia}
\affil{$^{\MWAaffil{CSIRO}}$CSIRO Astronomy and Space Science, PO Box 1130, Bentley WA 6102, Australia}
\affil{$^{\MWAaffil{WSU}}$Western Sydney University, Locked Bag 1797, Penrith, NSW 2751, Australia}
\affil{$^{\MWAaffil{CSIRO-NSW}}$CSIRO Astronomy \& Space Science, P.O. Box 76, Epping, NSW 1710, Australia}
\affil{$^{\MWAaffil{Haystack}}$MIT Haystack Observatory, Westford, MA, 01886-1299, USA}
\affil{$^{\MWAaffil{UToronto}}$Dunlap Institute for Astronomy and Astrophysics, University of Toronto, ON, M5S 3H4, Canada}
\affil{$^{\MWAaffil{ASTRON}}$ASTRON, the Netherlands Institute for Radio Astronomy, Postbus 2, NL-7990 AA Dwingeloo, the Netherlands}
\affil{$^{\MWAaffil{Melbourne}}$School of Physics, University of Melbourne, Parkville, Victoria 3010, Australia}
\affil{$^{\MWAaffil{Kumamoto}}$Graduate School of Science and Technology, Kumamoto University, Kumamoto, 860-8555, Japan}
\affil{$^{\MWAaffil{Calgary}}$University of Calgary, Calgary, AB, Canada T2N 1N4}
\affil{$^{\MWAaffil{NAOC}}$National Astronomical Observatories, CAS, Beijing 100012, China}
\affil{$^{\MWAaffil{Bologna}}$Dipartimento di Fisica e Astronomia, Universit\`a degli Studi di Bologna, via P. Gobetti 93/2, 40129 Bologna, Italy}
\affil{$^{\MWAaffil{INAF}}$INAF - Istituto di Radioastronomia, via P. Gobetti 101, 40129 Bologna, Italy}
\affil{$^{\MWAaffil{CurtinCIRA}}$Curtin Institute of Radio Astronomy, Curtin University, GPO Box U1987, Perth WA 6845}
\affil{$^{\MWAaffil{ASTRON2}}$ASTRON, Netherlands Institute for Radio Astronomy, Oude Hoogeveensedijk 4, 7991 PD, Dwingeloo, The Netherlands}
\affil{$^{\MWAaffil{UW}}$Department of Physics, University of Washington, Seattle, WA 98195, USA}
\affil{$^{\MWAaffil{CAASTRO}}$ARC Centre of Excellence for All-sky Astrophysics (CAASTRO)}
\affil{$^{\MWAaffil{INAF}}$Istituto Nazionale di Astrofisica (INAF) -- Istituto di Radio Astronomia, Via Piero Gobetti, Bologna, 40129, Italy}
}

\author[Beardsley et al.]{A.~P.~Beardsley$^{\MWAaffil{ASU}}$\thanks{email: Adam.Beardsley@asu.edu}\,,
M.~Johnston-Hollitt$^{\MWAaffil{Curtin}}$,
C.~M.~Trott$^{\MWAaffil{Curtin},\MWAaffil{ASTRO3D}}$,
J.~C.~Pober$^{\MWAaffil{Brown}}$,
J.~Morgan$^{\MWAaffil{Curtin}}$,
D.~Oberoi$^{\MWAaffil{Tata}}$,
D.~L.~Kaplan$^{\MWAaffil{UWisc}}$,
C.~R.~Lynch$^{\MWAaffil{Curtin},\MWAaffil{ASTRO3D}}$,
G.~E.~Anderson$^{\MWAaffil{Curtin}}$,
P.~I.~McCauley$^{\MWAaffil{USydney}}$,
S.~Croft$^{\MWAaffil{UCBerkeley}}$,
C.~W.~James$^{\MWAaffil{Curtin}}$,
O.~I.~Wong$^{\MWAaffil{UWA}}$,
C.~D.~Tremblay$^{\MWAaffil{CSIRO}}$,
R.~P.~Norris$^{\MWAaffil{WSU},\MWAaffil{CSIRO-NSW}}$,
I.~H.~Cairns$^{\MWAaffil{USydney}}$,
C.~J.~Lonsdale$^{\MWAaffil{Haystack}}$,
P.~J.~Hancock$^{\MWAaffil{Curtin}}$,
B.~M.~Gaensler$^{\MWAaffil{UToronto}}$,
N.~D.~R.~Bhat$^{\MWAaffil{Curtin}}$,
W.~Li$^{\MWAaffil{Brown}}$,
N.~Hurley-Walker$^{\MWAaffil{Curtin}}$,
J.~R.~Callingham$^{\MWAaffil{ASTRON}}$,
N.~Seymour$^{\MWAaffil{Curtin}}$,
S.~Yoshiura$^{\MWAaffil{Melbourne}}$,
R.~C.~Joseph$^{\MWAaffil{Curtin},\MWAaffil{ASTRO3D}}$,
K.~Takahashi$^{\MWAaffil{Kumamoto}}$,
M.~Sokolowski$^{\MWAaffil{Curtin}}$,
J.~C.~A.~Miller-Jones$^{\MWAaffil{Curtin}}$,
J.~V.~Chauhan$^{\MWAaffil{Curtin}}$,
I.~Boji{\v c}i{\'c}$^{\MWAaffil{WSU}}$,
M.~D.~Filipovi{\'c}$^{\MWAaffil{WSU}}$,
D.~Leahy$^{\MWAaffil{Calgary}}$,
H.~Su$^{\MWAaffil{Curtin}}$,
W.~W.~Tian$^{\MWAaffil{NAOC}}$,
S.~J.~McSweeney$^{\MWAaffil{Curtin}}$,
B.~W.~Meyers$^{\MWAaffil{Curtin}}$,
S.~Kitaeff$^{\MWAaffil{UWA},\MWAaffil{CSIRO}}$,
T.~Vernstrom$^{\MWAaffil{CSIRO}}$,
G.~G\"urkan$^{\MWAaffil{CSIRO}}$,
G.~Heald$^{\MWAaffil{CSIRO}}$,
M.~Xue$^{\MWAaffil{Curtin}}$,
\change{C.~J.~Riseley$^{\MWAaffil{CSIRO},\MWAaffil{Bologna},\MWAaffil{INAF}}$},
\change{S.~W.~Duchesne$^{\MWAaffil{Curtin}}$},
J.~D.~Bowman$^{\MWAaffil{ASU}}$,
D.~C.~Jacobs$^{\MWAaffil{ASU}}$,
\builders{}\\
\affils{}
}

\jid{PASA}
\doi{10.1017/pas.\the\year.xxx}
\jyear{\the\year}

\usepackage{aas_macros}
\usepackage{hyperref} 
\hypersetup{colorlinks,citecolor=blue,linkcolor=blue,urlcolor=blue}

\hypersetup{draft}

\begin{document}

\begin{frontmatter}
\maketitle

\begin{abstract}
The Murchison Widefield Array (MWA) is an open access telescope dedicated to studying the low frequency (\SIrange{80}{300}{\mega\hertz}) southern sky. Since beginning operations in mid 2013, the MWA has opened a new observational window in the southern hemisphere enabling many science areas.
The driving science objectives of the original design were to observe \SI{21}{\centi\meter} radiation from the Epoch of Reionisation (EoR), explore the radio time domain, perform Galactic and extragalactic surveys, and monitor solar, heliospheric, and ionospheric phenomena. All together 60$+$ programs recorded 20,000 hours producing 146 papers to date. 
In 2016 the telescope underwent a major upgrade resulting in alternating compact and extended configurations. Other upgrades, including digital back-ends and a rapid-response triggering system, have been developed since the original array was commissioned. 
In this paper we review the major results from the prior operation of the MWA, and then discuss the new science paths enabled by the improved capabilities. We group these science opportunities by the four original science themes, but also include ideas for directions outside these categories.

\end{abstract}

\end{frontmatter}

\section{INTRODUCTION }
\label{sec:intro}

The Murchison Widefield Array (MWA) is a low-frequency (80 -- 300 MHz) radio interferometer located at the Murchison Radio-astronomy Observatory (MRO) in Western Australia, the site of the future low band Square Kilometre Array (SKA-Low). 
The array was built with four primary science goals (for a full review, see \citealt{Bowman:2013}): detection of the Epoch of Reionisation (EoR); galactic and extragalactic (GEG) science; time domain astrophysics; and solar, heliospheric, and ionospheric  (SHI) science. These science drivers motivated a flexible design with 128 electronically beamformed, large field of view (FoV; 25 degrees at 150 MHz) antenna tiles, a densely packed core, and smooth \emph{uv} coverage to nearly 3km \citep{Tingay:2013}.
The MWA has been operational since 2013, and data from the telescope has been used to make progress along each of its science themes. 

The EoR collaboration has studied foregrounds and systematics extensively \citep[e.g.][]{line17, carroll16, procopio17, thyagarajan15, offringa16, jordan17, trott18}, and placed competitive limits on the redshifted 21cm power spectrum \citep{beardsley16, trottchips2016,ewallwice16,dillon15}.

The GEG group has produced the GaLactic and Extragalactic All-sky MWA (GLEAM) catalog of over 300,000 radio sources with declination south of $+30^{\circ}$ \citep{GLEAM, hw17}, successfully mapped 306 HII regions in the Galaxy \citep{2016PASA...33...20H}, detected molecules below 700 MHz \citep[e.g.][]{Trem+17}, placed limits on the surface brightness of the synchrotron cosmic web \citep{2017MNRAS.467.4914V},
and mapped the polarized diffuse sky at long wavelengths \citep{Lenc2016}.

The MWA has been used to follow up gravitational wave events \citep{ligo16} and gamma-ray bursts \citep{2015ApJ...814L..25K}, place limits on low-frequency Fast Radio Bursts \citep[FRBs;][]{2015AJ....150..199T, 2016MNRAS.458.3506R, 2016Natur.530..453K, msok_mwa_askap}, 
detect polarized flares from UV Ceti \citep{Lynch2017b},
survey the southern sky for low-frequency variability \citep{Bell:2018},
and perform detailed pulsar studies \citep[e.g.][]{2014ApJ...791L..32B,2016ApJ...818...86B,McSweeney:2017,2017ApJ...851...20M,xue2017, Bell:2016, Murphy:2017}.

Solar observations require extremely high dynamic range, which has led to new imaging and calibration techniques \citep{Oberoi17,Mohan17}. The SHI group has characterized weak non-thermal solar emission \citep{Suresh17,Sharma18}, and detected interplanetary scintillation due to solar wind both serendipitously \citep{Kaplan15} and with directed observations \citep{Morgan18}. One of the most exciting discoveries from the MWA has been the first direct detection of plasma ducts aligned with the Earth's magnetic field in the ionosphere  \citep{Loi2015a}. 

In addition to being a scientifically flexible instrument, the relatively simple front end infrastructure and compute based backend of the MWA make it a strong candidate for continued investment with upgrades that greatly enhance the capabilities at low cost. 
Since the initial deployment, the MWA has undergone several upgrades, including the addition of 128 tiles resulting in two distinct configurations (\emph{compact} and \emph{extended}) \citep{Wayth:2018}, an improved triggering system \citep{fast_response}, and a new digital back-end to support the Search for Extra-Terrestrial Intelligence (SETI, Croft et al. in prep.).
These upgrades have grown the scientific capabilities of the MWA, and in some cases enable entirely new directions to explore.

The MWA Collaboration currently includes 21 partner institutions from six countries (Australia, Canada, China, Japan, New Zealand, and the United States).
The MWA operates under an Open Skies policy, and any researcher may propose for observing time\footnote{Calls for proposals are advertised biannually on the MWA public website, \url{http://www.mwatelescope.org/}}. 
In addition, all raw visibility data becomes open access after an initial proprietary period and can be accessed by the All-Sky Virtual Observatory\footnote{\url{https://asvo.mwatelescope.org}}.

In this paper we give a brief overview of changes that have been made to the MWA since its initial deployment (Section~\ref{sec:upgrades}), and highlight many science opportunities that are enabled by these upgrades, categorized by the four primary science themes that drove the instrument design (Sections~\ref{sec:eor}-\ref{sec:shi}). 
The flexibility of observatory and its general-purpose nature have been key to its success. We recognize not all results will fall neatly within the four science themes, and present some examples of such opportunities in Section~\ref{sec:other}.
\section{UPGRADES AND IMPROVEMENTS}
\label{sec:upgrades}

We will refer to the MWA as described in Tingay et al. (2013) and operating 2013--2016 as ``Phase~I.'' Beginning mid 2016, 128 tiles were added to the array. We will refer to the array at this stage, along with other upgrades described below as ``Phase~II.''

\subsection{Additional tiles}
The antenna layout of the MWA was substantially changed with the addition of 128 tiles, for a total of 256 tiles. 
This upgrade is described fully in \citet{Wayth:2018}, and we provide a brief summary here. 
The 256-input correlator can process 128 dual-polarization signals at a time, so the array is periodically reconfigured between compact and extended configurations.

\subsubsection{Compact Configuration}
\begin{figure}
{
\includegraphics[width=\columnwidth,angle=0]{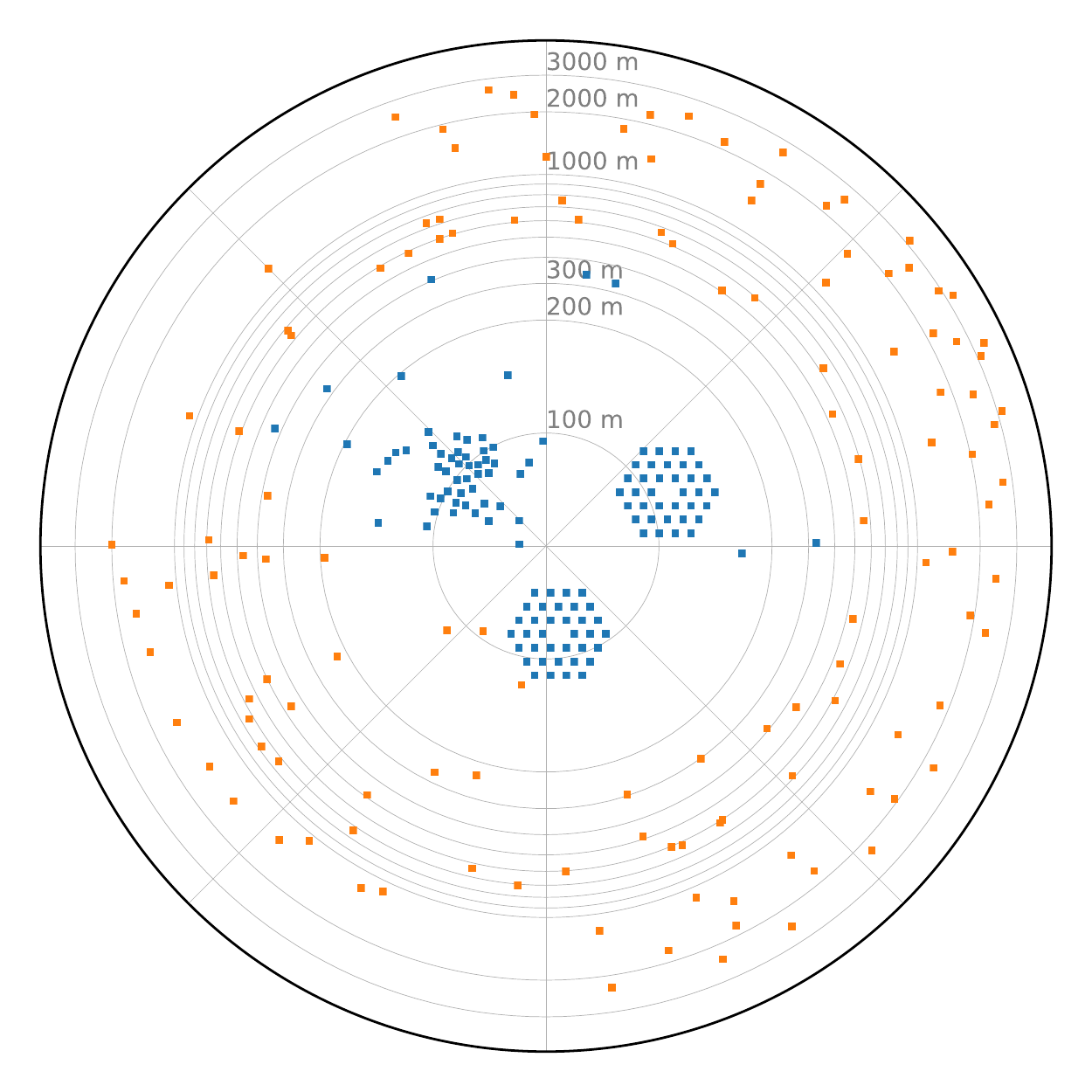}
}
\caption{The compact and extended configurations of the Phase~II MWA. The blue and orange squares show the tiles which are correlated in the compact and extended configurations respectively. Note the linear radial scale within 200~m to show the dense pseudo-random/redundant hybrid core, and logarithmic radial scale beyond to capture the nearly 6~km diameter.}
\label{fig:layout}
\end{figure}

The compact configuration of Phase~II consists of 56 tiles from the original Phase~I core (including a few tiles an intermediate distance from the center) and adds two hexagonal (``hex'') cores of 36 tiles each. The size of the hex cores were chosen to be comparable to the original core, and the spacing was chosen to yield relatively smooth \emph{uv} coverage to $\sim$~100 m. The spacing is also approximately that of the Hydrogen Epoch of Reionization Array \citep[HERA;][]{deboer17}, which will allow cross checks for EoR science. The layout is shown in Figure~\ref{fig:layout}.

\begin{figure}
\includegraphics[width=\linewidth]{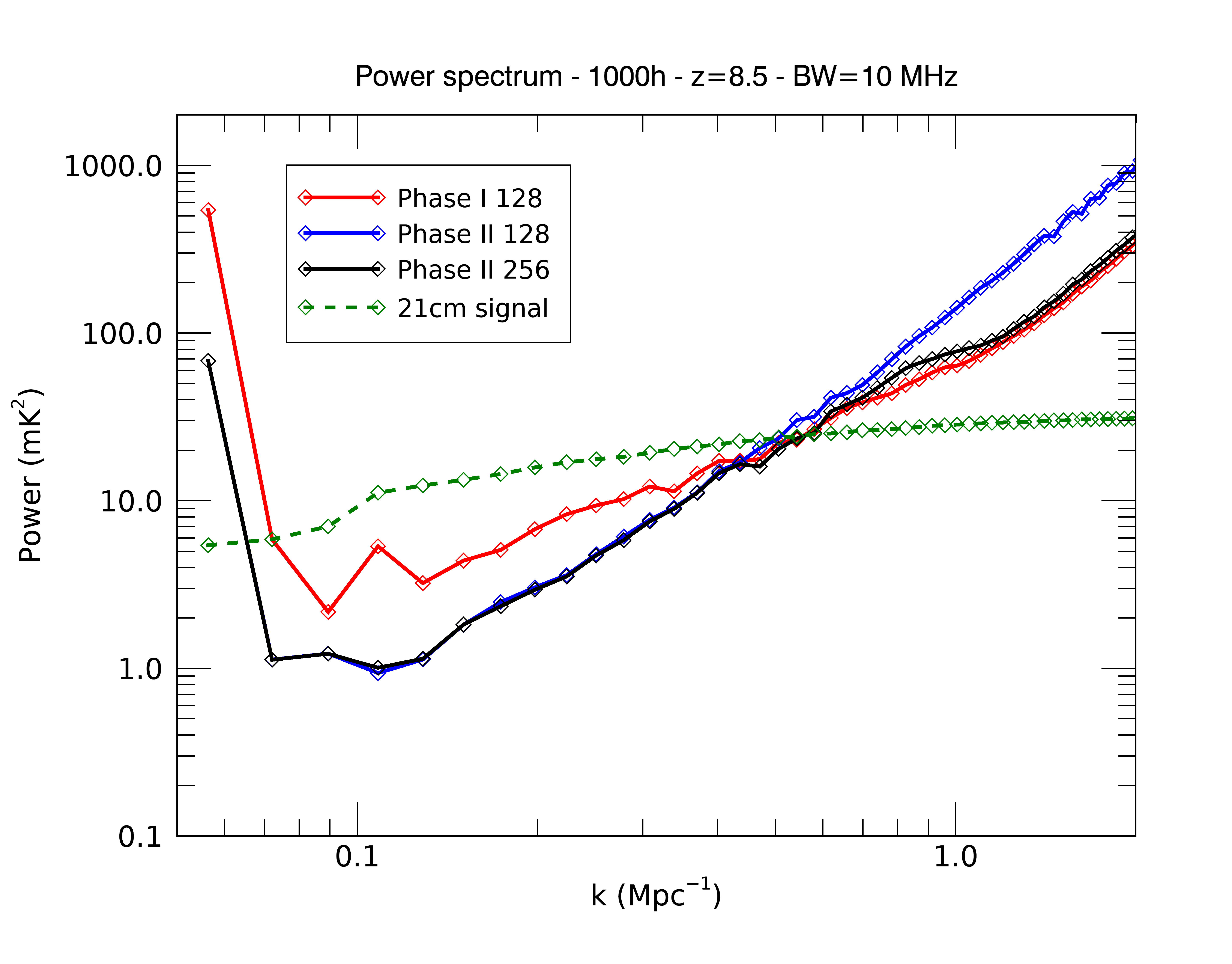}
\caption{Typical EoR power spectrum model at 150\,MHz with associated noise levels available to the Phase~I and Phase~II arrays with 1000\,h observation. ``Phase~II 256'' represents the result from a future MWA upgrade where all 256 tiles are used simultaneously. (From Wayth et al. 2018)}
\label{fig:eor_snr}
\end{figure}

The primary motivation for the compact configuration is to increase surface brightness sensitivity relative to the Phase~I array.  The net effect is a significant increase in sensitivity to diffuse signals like the 21\,cm signal from the EoR.  Figure \ref{fig:eor_snr} demonstrates the thermal noise levels achievable in a power spectrum measurement of the 21\,cm signal for Phase~I (red), Phase~II (blue), and a hypothetical Phase~II array with all 256 tiles correlated (black).  Phase~II achieves lower noise levels by a factor of $\sim 2$ to 5 (in mK$^2$) compared with Phase~I over a wide range of scales.
\change{Comparisons with other facilities such as HERA, the Low-Frequency Array (LOFAR), and the Precision Array for Probing the Epoch of Reionization (PAPER) can be found in \citet{deboer17} (their Figure 4) and the appendix of \citet{pober14}.}

Beyond the increased surface brightness sensitivity, the substantial number of redundant baselines serve a dual-purpose.  First, they enable coherent addition of redundant visibilities without requiring gridding or imaging.  This redundancy can ultimately serve to increase the sensitivity of a class of non-imaging EoR power spectrum estimators like the delay spectrum approach pioneered by \change{PAPER} \citep{parsons12a, parsons12b}
For a pedagogical description of imaging and non-imaging EoR power spectrum estimators, see \citet{Morales:2019}. 
Second, redundancy enables a new class of interferometric calibration techniques that derive antenna-based gains by minimizing deviations between visibilities from redundant baselines \citep{wieringa92,liu10,zheng14}.  The Phase~II compact configuration is somewhat unique, however, in retaining a large number of non-redundant baselines to preserve a high-fidelity point spread function for imaging and image-based calibration.

\subsubsection{Extended Configuration}
The extended configuration of Phase~II consists of 72 tiles from the original Phase~I array and 56 new long baseline tiles (Fig.~\ref{fig:layout}). The extended layout nearly doubles the longest baselines (from 3~km to 5.3~km), resulting in a resolution of $\sim 1.3'$ at 154\,MHz. \citet{Franzen:2016} estimated the Phase~I classical confusion limit to be $\sim$1.7~mJy at 154~MHz, and the improved resolution is expected to reduce this by a factor of 5 -- 10.

In addition, the naturally-weighted point spread function of the extended configuration is significantly smoother than that of the Phase~I array, owing to the lack of a dense core and instead a more uniform sampling of the \emph{uv} plane \citep{Wayth:2018}. Therefore we expect lower sidelobe confusion noise, and improved point source sensitivity.

\subsection{Digital Back-ends}
\label{sec:digital_backends}

\change{
\subsubsection{Correlator Data Rate}
The data rate of the archive restricted the correlator during Phase~I to a time-frequency resolution set by $ \tau \Delta f  \gtrsim 2\,000$, where $\tau$ is the visibility integration time and $\Delta f$ is the bandwidth of a single frequency channel.
However, the longer baselines of the extended configuration of Phase~II necessitated finer time-frequency resolution.
This was made possible by implementing a lossless in-situ compression on the visibilities \citep{Kitaeff:2015} and an expansion of the onsite archive system.
The result is a factor of four increase in possible output data rate, for example typical observations in the extended configuration have frequency and time resolution of \SI{10}{\kilo \hertz} and \SI{0.5}{\second}, respectively.
}

\subsubsection{Voltage Capture Buffer}
The MWA Voltage Capture System \citep[VCS;][]{2015PASA...32....5T} facilitates recording of raw antenna voltage data (before correlation and tied-array beamforming) at \SI{100}{\micro\second} and \SI{10}{\kilo\hertz} resolution. 
 This capability has been exploited for a variety of pulsar science applications such as investigating pulsar emission physics and studies of millisecond pulsars \citep[e.g.][]{2016ApJ...818...86B,McSweeney:2017,2017ApJ...851...20M,2018ApJS..238....1B,2018ApJ...869..134M}.
A new feature has enabled buffering voltage data up to \SI{150}{\second}, which can be recorded after receiving a trigger (\autoref{sec:rapid}). 
In this mode, instead of recording directly to disk, the voltages are stored in a ring buffer in the on-board memory of the VCS servers \citep{fast_response}. The voltages are kept in the rolling memory buffer for as long as possible (depending on the available memory) on a first-in-first-out basis.

\subsubsection{Breakthrough Listen}
Although replacement of the correlator hardware was not included in the Phase~II upgrades, work is currently underway (Morrison et al., in prep.) to enable much higher frequency resolution and more flexible beamforming. Additionally, a fiber link from the MRO to Curtin University means that raw voltage data can be accessed offsite, enabling easier deployment of commensal instruments that can produce beams and spectra within the primary field of view, independent of the science user who is controlling the primary beam pointing. The Breakthrough Listen \citep[BL;][]{2017Wordenetal} team has deployed hardware to Curtin as part of a pilot program (Croft et al., in prep) to perform experiments to search for extraterrestrial intelligence (SETI; Section~\ref{sec:seti}). The BL hardware will also serve as a general purpose instrument for targets such as fast transients and pulsars --- in essence an enhanced version of the existing VCS.

\subsection{Rapid-response Triggering}
\label{sec:rapid}
Recently, an upgraded automatic observation triggering system has been deployed on the MWA.
\cite{fast_response} describe the triggering system in detail, while we provide a brief overview here. The system handles alerts from the Virtual Observatory Event standard \citep[VOEvent;][]{Seaman_ivoa_2011} and can interrupt ongoing observations based on project priorities set ahead of time by the MWA director. Observations can be triggered in one of three modes: 1) using the regular visibility correlator; 2) using the VCS to capture voltage data; or 3) if the telescope is already in a buffered capture mode, the trigger can cause the buffer to be drained to disk. Due to scheduling constraints and processing time, the first two modes have a latency of \SIrange{6}{14}{\second} from the time the alert is received, while the third mode can have an effective negative latency because it can hold up to \SI{150}{\second} of buffered data.

\vspace{3ex}
\noindent
The remaining sections of this article will recap some of the progress to date in the MWA's key science themes and discuss further science opportunities enabled by the Phase~II upgrades. 
\section{EPOCH OF REIONISATION}
\label{sec:eor}

\begin{figure}
{
\includegraphics[width=\columnwidth,angle=0]{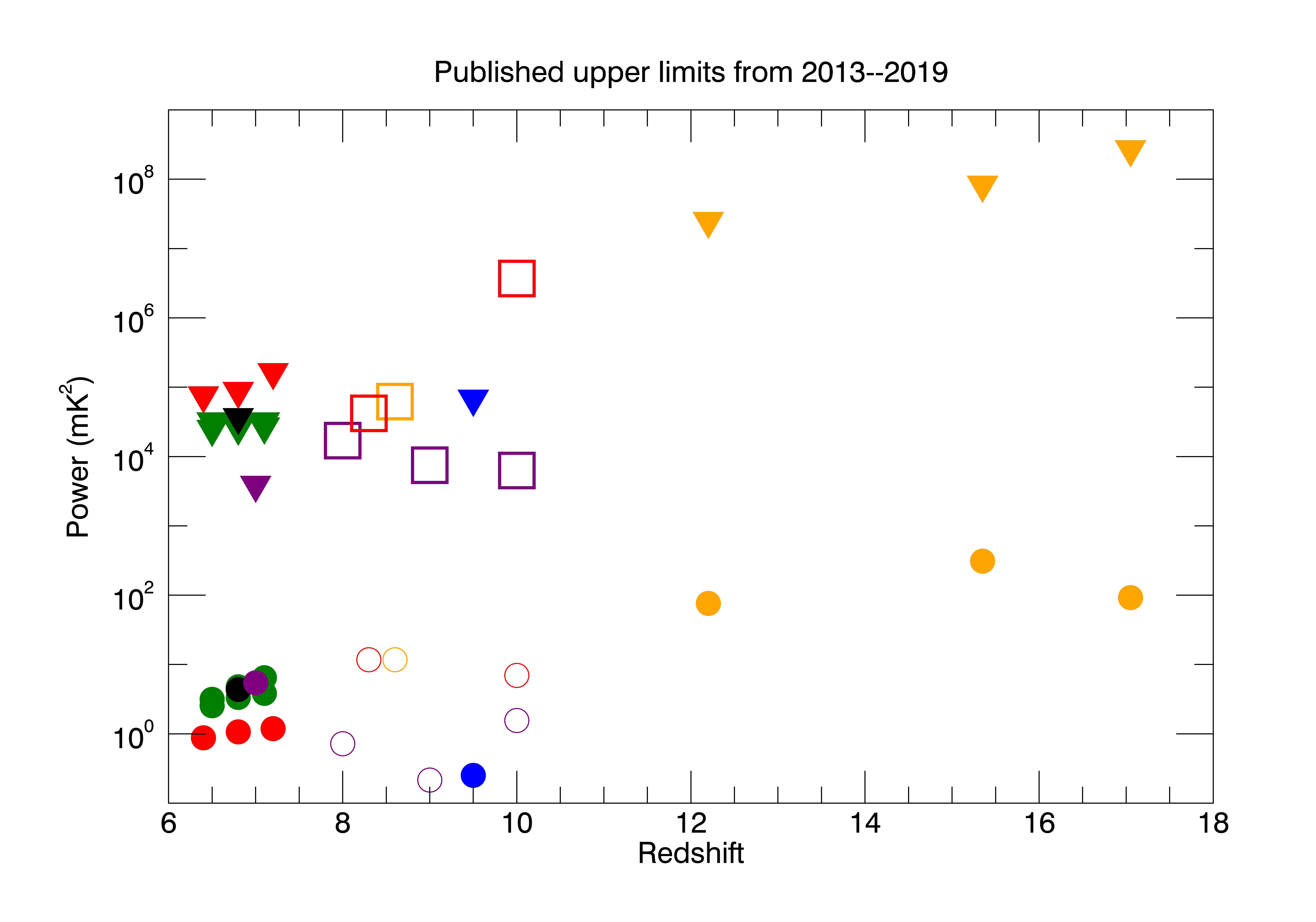}
}
\caption{Upper limits (95\% confidence) on the power spectrum of brightness temperature fluctuations from the Epoch of Reionisation at their respective redshifts and $k$-modes, published using MWA Phase I data (\change{filled} triangles): 
\change{\citet{barry19} (purple),}
\citet{trottchips2016} (red), 
\citet{beardsley16} (green), 
\citet{dillon15} (black), 
\citet{Dillon:2014} (blue), \change{and}
\citet{ewallwice16} (orange)\change{.}
\change{Leading results from other telescopes are shown with unfilled squares:
\citet{paciga13} (GMRT, orange),
\citet{patil17} (LOFAR, purple), and
\citet{kolopanis19} (PAPER, red).
Expected} signal strength using 21cmFAST for the same redshifts and scales \change{are shown with corresponding circles} \citep{mesinger11}.}
\label{fig:eor_limits}
\end{figure}

The MWA Epoch of Reionisation project collected more than 2000 hours of data during the four years of Phase I. These data were observed in pointed and zenith drift mode, and concentrated primarily on three southern fields away from the Galactic plane. \citet{jacobs16} describes the observational parameters and data analysis pipelines. Subsets of these data were published to place upper limits on the EoR spatial brightness temperature fluctuation power spectrum at $z=7.0-8.6$ \citep{barry19,trottchips2016,beardsley16,paul16,ewallwice16,dillon15}, as shown graphically in Figure \ref{fig:eor_limits} together with expected signal strengths from 21cmFAST \citep{mesinger11}. A larger body of research used these data to design, test and improve data analysis methodology and calibration schemes \citep{barry16,line17,procopio17,carroll16,offringa16}, and to explore aspects of the dataset and instrument that affect the potential to detect the EoR signal \citep{jordan17,Lenc+17,loi16,pober16}. The combined output of these analyses has demonstrated that exquisite knowledge of the sky, instrument and data processing methodology is crucial for successful and robust EoR detection. As such, instrumentation advances such as the hexagonal redundant sub-arrays of compact Phase II, combined with the long baselines and excellent snapshot $uv$-coverage of the extended configuration, afford new avenues for data calibration and sky model building.

\subsection{Redundant Calibration}

Precision calibration is one of the most significant challenges facing EoR experiments, as spurious spectral structure in calibration solutions can spread foreground power through $k$ space \citep[see e.g.][]{yatawatta15, thyagarajan15a, thyagarajan15b, barry16,dillon18}.  \cite{barry16}, in particular, demonstrate how calibration algorithms that rely on a sky model can introduce contamination into the EoR window when the sky model is incomplete.  Redundant calibration \citep{wieringa92,liu10,zheng14} therefore presents an appealing alternative for EoR experiments, because it only requires a sky model to constrain a subset of the calibration parameters.  Because of its unique layout, with substantial numbers of both redundant and non-redundant baselines, the compact configuration of Phase II is a valuable testing ground for comparing redundant and sky-based calibration techniques in the pursuit of EoR science.  \cite{li18} conducted just such a study, developing redundant calibration techniques for the Phase II compact array, and comparing them with the sky-based calibration techniques similar to those used in \cite{beardsley16}.  They find that applying redundant and sky-based calibration techniques in tandem yield small but significant reductions in foreground contamination of the EoR power spectrum.

A number of other theoretical studies are under way to determine optimal strategies for calibrating data from the compact array for EoR science, augmenting existing literature of results from
other arrays \citep{thyagarajan18, orosz18}.  \citet{joseph:2018} demonstrate how the flux density distribution of the sky can affect redundant calibration techniques, despite the lack of explicit reference to a sky model.  \citet{Byrne:2019} study the subset of parameters that cannot be constrained by redundant calibration and show how sky-model incompleteness errors still affect these terms, which, in turn, introduce contamination into the EoR window.  Continued studies of redundant and sky-based calibration techniques using the compact array may be invaluable for the future of 21\,cm science, as a combination of both techniques may be necessary to mitigate the limitations of each one independently.

\subsection{21\,cm Power Spectra}

As stated in Section \ref{sec:upgrades}, the compact configuration of Phase II provides a significant increase in surface brightness sensitivity over the Phase I configuration.  Figure \ref{fig:eor_snr} shows how this upgrade translates into expected improvements in the sensitivity of the array to the power spectrum of the EoR.  \cite{li18} present the first preliminary power spectra from Phase II, using a few hours of data, and show that existing power spectrum pipelines from Phase I like FHD/$\epsilon$ppsilon \citep{jacobs16} can be adapted and applied to Phase II data.  Work is now in progress to process the first season of Phase II data and produce a deep power spectrum limit comparable to \cite{beardsley16}.  Other efforts are under way to analyse data taken from Phase II in a drift scan mode using alternate power spectrum pipelines.

\subsection{LoBES and Diffuse Emission}
\label{sec:lobes}
The MWA EoR fields were chosen based on their low sky temperature and preferable elevation angles for night-time observing. Yet these fields are imperfect: the large field of view of the MWA, extending beyond 40$^{\circ}$ in the sidelobes, makes it difficult to avoid all bright extended radio galaxies and the Galactic plane entirely. Therefore the MWA EoR fields contain several bright, extended sources (e.g. Hydra A, Fornax A, 3C444) located either at the edges of the MWA primary beam or in the primary beam sidelobes \citep{jacobs16}. These sources are expected to produce significant foreground contamination in the EoR power spectrum, with \citet{procopio17}
estimating that mis-modelling the brightest ten sources contributes $>$90\% of the power
bias. \citet{trottwayth17} compared bias in the EoR power spectrum from poorly-resolved
extended sources, showing that detection of the EoR with the MWA would require the
spatial resolution of the extended MWA combined with some TGSS data from the Giant Metrewave Radio Telescope (GMRT).

The frequency-dependence of an interferometer's point spread function (PSF) becomes stronger far from the pointing centre and so sources located far from the primary field of view produce more foreground contamination than sources located in the centre of the primary beam \citep{thyagarajan15,trott12}. \citet{pober16} showed that subtracting a foreground model that includes sources in both the main field of view and the first sidelobe is found to reduce the contamination in EoR power spectrum by several per cent relative to a model including only the sources in the main field of view, and other recent work has studied the impact of incomplete sky models on calibration \citep{patil16,barry16}. Additionally, detailed models for extended sources and double-sources are required for foreground modelling and subtraction, as subtracting these sources as point sources leaves residual excess power that has the potential to introduce bias into the EoR power spectrum \citep{procopio17,trottwayth17}.

\begin{figure}
\includegraphics[width=.97\columnwidth]{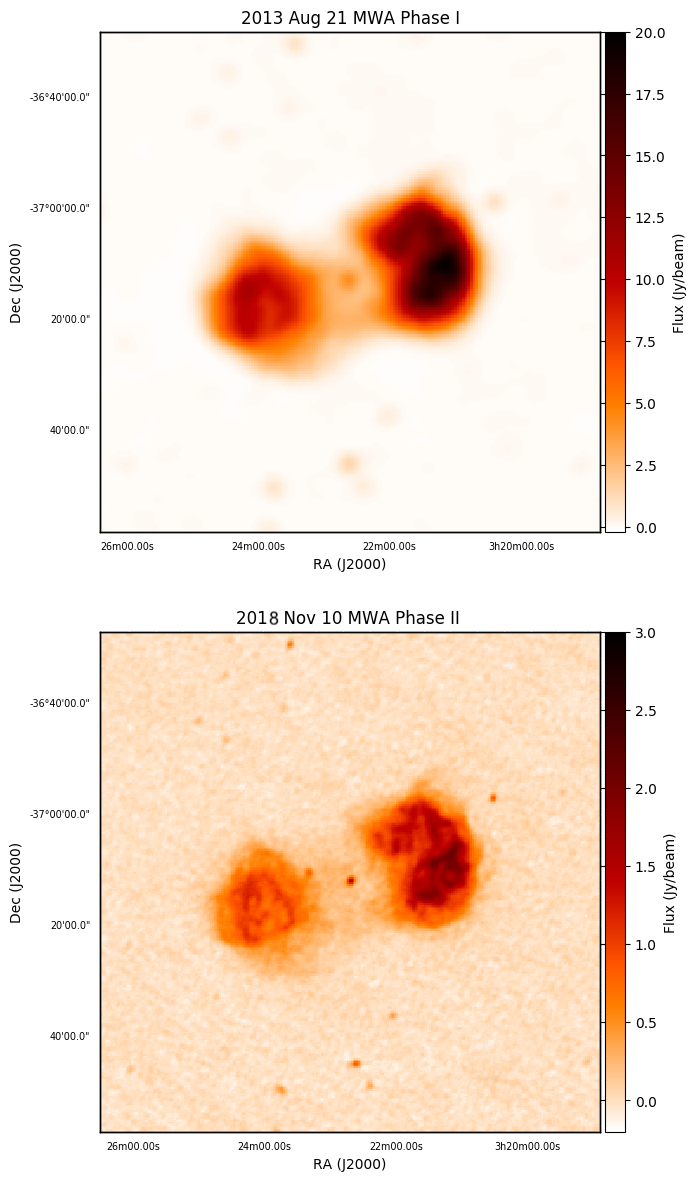}
\caption{Comparison of MWA Phase I and Phase II extended array images of Fornax A. The extended array resolves the finer structures in the lobes of this source while over-resolving the bright, more diffuse emission. The MWA Phase~I image adapted from \citet{mckinley15}.}\label{fig:fornax}
\end{figure}

In an effort to improve the source models of both point and extended sources in the MWA primary beam sidelobes of the EoR0 (Right Ascension (RA) = \SI{0.00}{\hour}, Declination (Dec) = \SI{-27}{\degree}) and EoR1 (RA = \SI{4.00}{\hour}, Dec = \SI{-27}{\degree}) fields, 
members of the MWA EoR team are conducting  the Long Baseline Epoch of Reionisation Survey (LoBES). This survey consists of multi-frequency (four bands covering 103\ --\ 230\ MHz) observations of EoR0, EoR1, and their eight neighbouring fields using the MWA Phase II extended array. Observations were undertaken in Semester 2017B and are being calibrated for future publication (C.~R. Lynch, in preparation). Observations of the EoR0 and EoR1 fields using the longer baselines will provide high-resolution $uv$-components to compliment the existing $uv$-plane sampling of these fields. The neighbouring fields will provide high-resolution observations of troublesome sidelobe sources within the centre of the MWA primary beam, where the primary beam is well modelled \citep{sutinjo15}. As an example, Figure \ref{fig:fornax} compares MWA Phase I and Phase II extended array images of Fornax A, highlighting the more complex structure revealed by the higher resolution of the extended array. The longer baselines will also allow for deeper measurements of foreground sources due to the reduced classical confusion limit, expanding the foreground catalogue by a factor of \numrange[range-phrase=\,--\,]{3}{4}. 
Using a model of extragalactic foreground contamination similar to that of \citet{trottchips2016}, this could reduce the amount of leaked foreground power by an estimated factor of \numrange[range-phrase=\,--\,]{e2}{e3}. 

\subsection{21cm-LAE crosspower spectrum}
The cross correlation between the 21~cm signal and the high redshift
Lyman-$\alpha$ emitter (LAE) distribution can help to identify the 21~cm signal and enhance the detectability
because the foregrounds are expected to have negligible correlation with the
LAE distribution \citep{2016MNRAS.459.2741S, Lidz_2008}. 
In the absence of foregrounds, the MWA Phase II compact configuration is predicted to have sufficient sensitivity to detect the 21~cm-LAE cross power spectrum (CPS) with an observation of 1000 hours combined with the LAE survey by Subaru Hyper Suprime Cam (HSC) \citep{Kubota2018}. With realistic models for foregrounds, \citealt{Yoshiura2018} found that subtraction of 80\% of diffuse and 99\% of point source foregrounds will be needed to reach a CPS detection at $k = 0.4$\,hMpc$^{-1}$. The removal of these contaminants is an ongoing effort by the MWA collaboration, and is aided by innovative Phase II surveys as described in Sections~\ref{sec:lobes} and \ref{sec:diffuse_fg}.

\begin{figure}
{
\includegraphics[width=0.5\textwidth,angle=0]{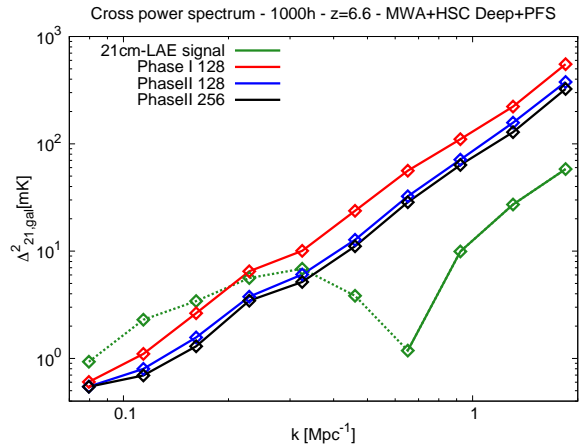}
}
\caption{Expected sensitivity to the 21cm-LAE power spectrum at $z$=6.6 with an observation of 1000h, HSC and PFS survey. The signal is calculated by using large scale radiative transfer simulation of reionisation, which is identical to the simulation used in \cite{Kubota2018}.}
\label{fig:eor_21cmLAE}
\end{figure}

The LAE survey is ongoing and, recently, the LAE distribution with partial survey data at $z$ = 5.7 and 6.6 has been reported \citep{Ouchi2018, Shibuya2018a, Shibuya2018b, Konno2018}. Thus, there is an opportunity for analysis of 21cm-LAE cross correlation. Although the areas of sky do
not overlap with the MWA EoR fields, the MWA Phase II will observe the HSC fields and can place a competitive limit on the CPS with a relatively short duration of survey
because the Compact Phase II has high sensitivity at large scales as shown in Figure \ref{fig:eor_snr}. Furthermore, the detectability is
enhanced when a spectrographic survey by the Prime Focus Spectrograph (PFS)
determines the precise redshift of LAEs, and the 3-dimensional information is available. The expected sensitivity is shown in Figure \ref{fig:eor_21cmLAE}.

\subsection{21cm Bispectrum}
The bispectrum is the Fourier Transform of the three-point correlation function, and extracts non-Gaussian information about the 21cm brightness temperature field that is lost in two-point statistics such as the power spectrum \citep{watkinson17,watkinson18,mondal15,majumdar18,yoshiura15}. Bispectra are fundamentally formed from closed triplets of interferometric baselines, with equilateral, isosceles and more generalised triangle configurations encoding information on different scales. The 21~cm bispectrum is difficult to measure with current experiments \citep{yoshiura15}, but allows a fresh approach to treating foregrounds by using the data in a different way to power spectra. In particular, the redundant triangles available in the hexagonal subarrays of the Compact Phase II configuration afford a direct bispectrum estimator with increased instantaneous sensitivity. In \citet{trott18}, a direct bispectrum estimator is applied to 20 hours of Phase II data, and compared with the estimates using a gridded bispectrum estimator, which uses visibilities gridded to the $uv$-plane to extract triangles. For stretched isosceles triangles that probe regions of parameter space outside of the wedge foreground contamination region, bispectral estimates are consistent with noise at the 20 hour integration stage. 

\subsection{Large Scale Foreground Mapping}
\label{sec:diffuse_fg}

With a compact core and massively redundant sampling of short baselines, the MWA Phase II is optimized for 
measurements of the power spectrum on large (many degree) scales. This is in large part because simulations of the 21cm 
signal generally suggest that the brightest modes occur on degree scales.  Expressed in units of \si{\milli\kelvin\squared \mega\parsec^{3}}, the predicted power spectrum, $P(k)$, rises as a power law with decreasing 
$k$, while noise remains flat \citep{mesinger11}. Unfortunately, a similar spectrum holds for smooth foreground emissions; at degree 
scales bright galactic power begins to dominate over extragalactic sources \citep[see e.g. ][]{beardsley16}. Accuracy of foreground models used in calibration and foreground subtraction continue to be a limiting factor in 21cm experiments both interferometric \citep{barry16} and global \citep{Mozdzen2019}; better maps are needed in the southern hemisphere. However, reconstructing the largest scales probed by an interferometer is difficult where deconvolution must distinguish between primary beam and true structure \citep{RauandCornwell2011}. Surveys by the Long Wavelength Array (LWA) and the Owens Valley LWA dipole arrays image the largest scales across the visible sky but do not cover the southern hemisphere \citep{Eastwood2018, Dowell2015}.



Reconstruction of large scale structure with the MWA could be pursued in two ways: 
mosaicing with many pointings and reconfiguring tile beamforming to widen the field of view.  Phase I EoR 
observations specifically targeted large scale structure by increasing $uv$ coverage on pointings flanking the primary 
EoR fields. These observations are currently under study. A similar program with the MWA Phase II, taking advantage of 
redundancy to supplement calibration models, could potentially offer higher calibration dynamic range as well as a 
model better matched to Phase II EoR observations.   Another method for reconstruction of power on large scales 
is the m-mode formalism \citep[e.g. ][]{Shaw2014,Eastwood2018} which is particularly well suited for very large field of view observations where a full 24 
sidereal hours are available.  This is the goal of the MWA Phase II m-mode project which is currently running a 
series of observations with all but one dipole per tile disconnected, increasing the field of view to nearly the full sky.








\section{GALACTIC AND EXTRAGALACTIC SCIENCE}
\label{sec:geg}
The category of ``Galactic and Extragalactic'' (GEG) science includes a wide variety of targets ranging from molecular transitions in our own Galaxy to large scale structure in the Universe.
The commonality between these objectives lies in the imaging analysis and the need for high resolution, low noise sky maps.
The extended configuration of the Phase-II upgrade offers significantly improved resolution, lower sidelobe confusion limit, and improved sensitivity in uniform weighted images.
These factors will enhance the scientific return on many of the GEG science goals, and improve synergy with other surveys.

\subsection{Galactic continuum}

\subsubsection{Cosmic ray tomography}
Below $\sim150$\,MHz, most \textsc{Hii}~regions become optically thick, and therefore appear in absorption relative to the background Galactic synchrotron. When the distance to the \textsc{Hii} region is known, the difference between its surface brightness and a nearby (source-free) region gives a direct measurement of the integrated cosmic-ray electron emissivity between the \textsc{Hii} region and the edge of the Galactic plane. 
Phase~I of the MWA was successfully used by \cite{2016PASA...33...20H} to detect and measure 306~\textsc{Hii}~regions. \cite{2017MNRAS.465.3163S, 2018MNRAS.479.4041S} went on to use the distinct spectral signature of these regions to perform cosmic ray tomography of the Galactic Plane. 
Foreground measurements, between the \textsc{Hii} region and Earth, are more difficult, as an absolute measurement of the emissivity must be calculated, which is impossible for an interferometer which does not measure total power. However, \cite{2018MNRAS.479.4041S} spectrally scaled the 408\,MHz total power image of \cite{1981A&A...100..209H, 1982A&AS...47....1H} to fit the foreground emissivity, albeit with large errors. Their results showed an increase in emissivity toward the Galactic Centre and a decrease with galactocentric radius, consistent with other results in the literature.

The limitations of this work are due to two main factors: the low resolution of Phase~I, which reduces both the number of detectable \textsc{Hii} regions due to confusion noise, and the separability of \textsc{Hii} regions in complex areas; and the lack of distance estimates toward \textsc{Hii} regions, which are necessary in order to perform the tomography measurement. The improved resolution of Phase~II MWA will considerably improve the detectability and separability of \textsc{Hii} regions, revealing $\sim 2\times$ more, which, with distance estimates, could be used for tomography. Upcoming radio recombination line surveys such as the \textsc{Hii} Region Discovery Survey \citep[HRDS; ][]{2011ApJS..194...32A, 2012ApJ...759...96B, 2015ApJS..221...26A} and its Southern counterpart, SHRDS \citep{2017AJ....154...23B}, aim to find distances to the hundreds of \textsc{Hii} regions catalogued by the \textit{Wide-Field Infrared Survey Explorer} \citep{2014ApJS..212....1A}. The combination of improved resolution and distance estimates also gives the ability to measure more distant (smaller apparent size) \textsc{Hii} regions, which yields more 3D sampling of the Galactic plane, considerably improving the leverage of the data over the models of cosmic ray electron distribution \citep{2011A&A...534A..54S} and magnetic field distribution \citep{2017ARA&A..55..111H}.

\subsubsection{Planetary Nebulae}
More than 300 planetary nebulae (PNe), visible to MWA, have angular sizes larger than 1~arcmin \citep{2006MNRAS.373...79P, 2009MNRAS.399..769F, 2008MNRAS.384..525M, 2016JPhCS.728c2008P}, which can be resolved by Phase~II of the MWA. Nearly all PNe are optically thick and will only show self-absorption features at the MWA frequencies, thus can be used to perform cosmic ray tomography, similarly with that of the \textsc{Hii} region absorption. Furthermore, excellent MWA low frequency coverage, in combination with high frequency measurements, will be extremely useful for PNe spectral energy distribution (SED) construction and examination of potential radial density gradients and \change{the} emission measure to angular diameter relationship \citep[][{Boji{\v c}i{\'c}} et al. in prep.;]{2017MNRAS.468.1794L}.

\subsubsection{Supernova remnants and Pulsar wind nebulae}

The currently detected population of supernova remnants (SNRs) is considerably lower than the total number of SNRs expected in our Galaxy \change{\citep{Modjaz2019}}. In particular, examination of pulsar creation rates, heavy element abundances, OB star counts and stellar life-cycle models, as well as detection rates of SNRs in other galaxies, all suggest that there should be well over 1000~SNRs detectable in the plane of the Galaxy \citep{1991ApJ...378...93L,1994ApJS...92..487T,2006ApJ...639L..25B}. To date only a fraction of these ($<30$~per~cent) have been detected despite the ever increasing number of observational surveys of the Galactic plane; over 90\% of those detections have been made at radio frequencies.

Older SNRs are expected to have various angular scales and low surface brightnesses \citep[e.g. G\,$108.2-0.6$, a low surface brightness SNR detected in the Canadian Galactic Plane Survey; ][]{2007A&A...465..907T}. It is detecting this latter category of SNRs to which the MWA is well-suited, with its superb diffuse source sensitivity. As expected \citep{Bowman:2013}, the Phase~I MWA proved itself to be a powerful machine for the detection of new SNRs \change{\citep[Hurley-Walker et al. in prep; Johnston-Hollitt et al. in prep]{Maxted:2019, Onic:2019}}, including several that had been previously misclassified as \textsc{Hii} regions \citep{2016PASA...33...20H}. These reclassifications were possible thanks to the low frequency coverage and the high spectral resolution of the MWA: the spectra of \textsc{Hii} regions turn over and go into absorption at the lowest part of the MWA band, and are thus readily distinguished from SNRs, which display power law spectra.

\begin{figure*}
    \centering
    \includegraphics[width=\textwidth]{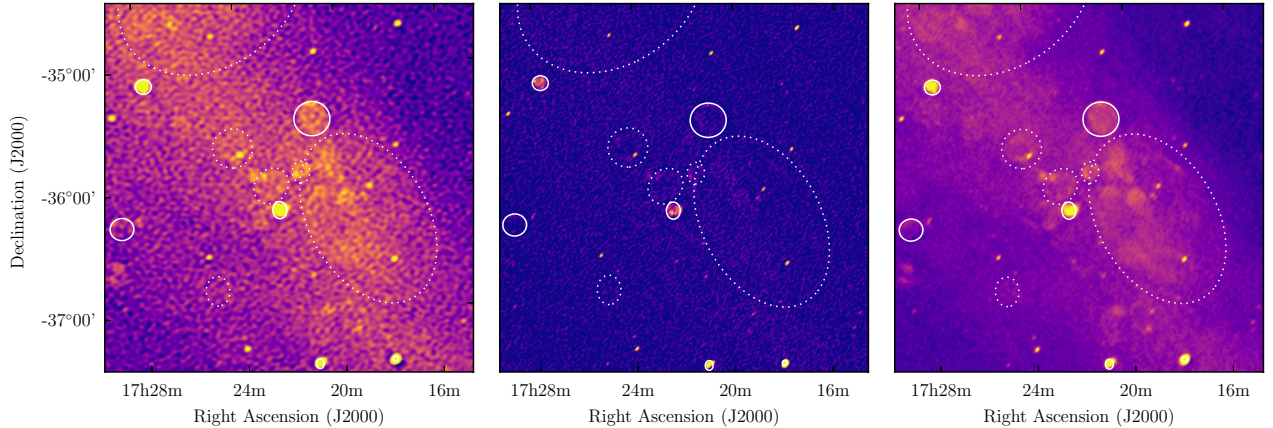}
    \caption{\change{A region of sky centred on RA 17$^h$22$^m$, Dec -36$^\circ$ at 139--170\,MHz is shown using three different datasets and imaging techniques. Left: A single two-minute snapshot from the original Phase~I configuration, imaged with multiscale \textsc{WSCLean} and a Briggs weighting of -1 (Hurley-Walker et al. 2019a, accepted); Middle: A single two-minute snapshot from the extended array, imaged with multiscale \textsc{WSClean} and a Briggs weighting of 0; Right: The two observations imaged together using image-domain-gridding \citep{vanderTol:2018} in \textsc{WSClean} and a Briggs weighting of 0. Known SNRs are shown with solid lines, while SNRs detected by Hurley-Walker et al. (2019b, accepted) are shown with dotted lines. The increased resolution and imaging quality of Phase~II MWA make it possible to discern these SNRs in four minutes, instead of $\approx30$\,minutes.}}
    \label{fig:natashas_fig}
\end{figure*}

However, the angular resolution of the Phase~I array confined detections to SNRs that had angular sizes greater than $0.1^\circ$, leaving potential populations of smaller SNRs waiting for discovery with the Phase~II array. 
Combining Phase~I and Phase~II data will potentially reveal low surface brightness emission below previous confusion limits, both within and outside of the Galactic Plane, \change{as well as having the higher angular resolution of Phase~II with the large-angular scale sensitivity of Phase~I (see Figure~\ref{fig:natashas_fig} for an example of the combined imaging potential).}
The complete population of SNRs in the nearby Magellanic Clouds will also be detectible \citep{2017ApJS..230....2B,for18}.

\subsection{Nearby galaxies and Magellanic clouds}

Galaxy evolution is mainly governed by the combination of  physical
processes within the galaxy's interstellar medium (ISM), and
interactions between the galaxy \change{and} its local environment \change{\citep[e.g.,][]{Dickey2010}}. The MWA
Phase~II will enable significant progress within the field of galaxy
evolution through integrated low-frequency synchrotron spectra studies
of many more star-forming galaxies and active galactic nuclei (AGN) than was previously possible
due to the order of magnitude lower noise floor compared with MWA Phase~I.  The low-frequency radio properties of low-redshift
star-forming galaxies, for which much multi-wavelength data are
available, will now be accessible through direct detections and stacking
techniques.

Resolved radio observations of galactic disks, in combination with
observations at other wavelengths such as the far-infrared (FIR), relate the ISM
process of star formation to the magnetic fields that generate the
observed radio emission observed at low frequencies \citep[e.g.\
][]{schleicher16, 2006MNRAS.370..363H, 2013Ap&SS.343..301L}. \citet{klein18} show that the synchrotron spectra of
nearby star-forming galaxies is not well-represented by a simple power
law extending from MHz to GHz frequencies. Rather, the synchrotron
spectrum at low frequencies (below 1~GHz) is rather flat, with a
spectral index ($\alpha$ where $S_{\nu} \propto \nu^{\alpha}$) of
approximately $-0.6$,  coupled with a break and an exponential decline
at GHz frequencies. Previous studies show that the MWA Phase~I
synthesised beam is too large to allow for resolved radio spectral index
maps of  specific star-forming regions even in the nearest galaxies, such
as the Magellanic Clouds \citep{for18} and NGC~253
\citep{kapinska17}.

In the neighbouring Magellanic Clouds, the MWA Phase~II capabilities will
enable the quantification of the cosmic ray energy spectrum and the
measurement of spatial variations that arise from shock re-acceleration,
spectral aging, and absorption effects $-$ the primary goal of The Deep Survey of the Magellanic System (MAGE-X\change{, \citealt{for18}}).
The importance of hydrodynamical environmental processes such as ram pressure is understood in the cluster environment \citep[e.g.,][]{Kenney:2004} as well as galaxy groups with known hot IGM \citep[observed in X-rays; e.g., ][]{Rasmussen:2006}.  
\citet{Murphy:2009} also suggest that the intracluster wind can sweep the cosmic ray electrons and associate magnetic fields to one side, creating synchrotron tails.
However, the importance of these processes in low-density environments such as galaxy groups is less clear. 
Recent studies suggest that ram pressure may be occurring in lower-density galaxy groups without hot observable X-ray halos \citep{Westmeier:2013}.  
Ram pressure could also explain some of the extended radio halos that are observed in some but not all galaxies \citep[e.g.,][]{Heesen:2018}.
The role of ram pressure can be directly confirmed
through the detection of synchrotron emission from a high Mach number
shock at the eastern boundary of the Large Magellanic Cloud (LMC)
through the MAGE-X project.

Within the Magellanic Clouds, the combination of Phase~II observations
and soft X-ray emission (from the forthcoming eROSITA mission) will
enable the investigation of the SNR population and
the host ISM within the Clouds through detailed and global studies of
Magellanic Cloud SNRs and superbubbles \citep{2016A&A...585A.162M, 2019A&A...621A.138K}. Resolved analyses from the
MAGE-X project \change{are} complemented by multiwavelength ancillary
observations that include high-frequency radio observations
\citep{Kim03,Hughes07, 2012SerAJ.184...93W, 2011SerAJ.183..103W,2011SerAJ.183...95C,2011SerAJ.182...43W} in addition to forthcoming HI and continuum
observations from the Galactic Australian Square Kilometre Array Pathfinder \citep[GASKAP;][]{GASKAP} and Evolutionary Map of the Universe \citep[EMU;][]{EMU} surveys using ASKAP; the mid- and
far-infrared observations from Spitzer \citep[Surveying the Agents of Galaxy Evolution$-$SAGE; ][]{meixner06} and
Herschel \citep[The HERschel Inventory of The Agents of Galaxy Evolution$-$HERITAGE; ][]{meixner13}; and near-infrared observations
from the on-going VISTA survey \citep{cioni13,ivanov16}.  The Phase~II
observations will also have comparable angular resolution to large area
optical narrow band surveys, such as the Southern H-Alpha Sky Survey Atlas  \citep[SHASSA;][]{2001PASP..113.1326G}
or the Magellanic Cloud Emission Line Survey \citep[MCELS;][]{2012ApJ...755...40P} in H$\alpha$. 
Such H$\alpha$ observations can be used to separate the thermal from the
non-thermal components of the observed radio continuum observations
\citep[e.g.\ ][]{tabatabei07}. The resulting thermal and non-thermal radio continuum
maps can be used to derive ISM physical parameters, such as magnetic
field strength and thermal electron density of warm ionised gas.

Stellar winds and supernovae give rise to  \emph{bubbles} \citep{2014AJ....147..162D} and
\emph{superbubbles} \citep{2015A&A...573A..73K}.  The expansion of superbubbles is responsible for the
compression and fragmentation of cool gas, and is well-traced by
observations of HI, dust and molecules \citep{2017ApJ...843...61S}.  Such winds may also create
\emph{chimneys} which bridge a galaxy's star-forming disk and halo,
facilitating the propagation of cosmic rays towards the halo via the
magnetic fields in these chimneys \citep{norman89}. The abundance of extended radio halos,
which are observed in a few nearby
galaxies \citep[e.g.\ ][]{duric98,kepley10,srivastava14,kapinska17,2018A&A...611A..72K}, is
not well understood. Significant progress in this area will also further
our understanding of the connection between the galactic disk and the
halo -- currently one of the most active areas of investigation
\citep[e.g.\ ][]{bland17}. A more exotic theoretical model suggests that the observed excess of radio  continuum
emission in the halo of NGC\,1569 may be due to dark matter annihilation
\citep{ho18}. Therefore, the new observations, with improved sensitivity
and resolution, from the MWA Phase~II
will significantly further our understanding of the resolved ISM
processes that connect the galactic disks to halos, as well as provide
vital constraints to theoretical models.
\subsection{Spectroscopy}

One of the unexpected results from Phase~I of the MWA is the first detections of molecules below 700MHz \citep{Trem+17,Trem+18a}.  \citet{Trem+18b} also made tentative detections of carbon radio recombination lines around 106MHz.  However, Phase~II allows for better detection due to the reduced beam size which decreases the beam dilution.  A standard assumption when determining the column density or total intensity of the emission is that the source fills the telescope's synthesized beam.  With a \SIrange{2}{3}{\arcminute} beam in the surveys done with Phase~I, stellar sources at 400\,pc or greater distances, are likely significantly smaller than the beam, creating an error in the measurements.  With the longer baselines from Phase~II, the beam size decreases to about \SI{1}{\arcminute}, allowing for better detection of weak signals and signals from greater distances.  A future upgrade to the correlator will greatly improve the frequency resolution and therefore the sensitivity for currently unresolved spectral lines, further enhancing study of the kinematics and physical properties of interstellar gas for a vast range of astronomical objects.

\change{LOFAR \citep{vanHaarlem} has been the current leader in the study and analysis of low-frequency carbon recombination lines with studies of Cassiopeia A \citep{Asgekar,Salas}, Cygnus A \citep{Oonk14} and the extragalactic detection in M82 \citep{Morabito}.
Recently \cite{Salgado+17a,Salgado+17} published two in depth papers on the full theoretical analysis of carbon recombination lines at the quantum states of n--bound states $>$200 including the level population determination (influencing the strength of the detected lines) in order to develop carbon recombination lines as tools to study the physical conditions of the local gas.
Following on from the work using LOFAR in the northern hemisphere, the MWA can equivalently view the low-frequency sky from the southern hemisphere.
However, no other telescope has reported molecular line detections below \SI{700}{\mega \hertz}.}

There is a range of science cases for studying astronomical phenomena in molecular and atomic spectral lines at low frequencies: high-mass star formation \citep{Codella+14}, detection of complex organics to look for signs of life in stars and planets 
\citep{Danilovich+16}, reviving investigation of the disportionate ratio of organic and inorganic molecules \citep{Cosmovici+79}, and determining the physical properties of the interstellar medium \citep{Salgado+17}, including its cold diffused component \citep{Oonk+15,Peters+11}. The study of molecules around high-mass stars is complicated partially due to strong emission of prominent molecules which would not be present at lower radio frequencies.  In addition, spectroscopic capability at low frequencies offers the opportunity to search for highly redshifted (z > 5) neutral hydrogen absorption towards radio AGN in the early Universe \citep[e.g.][]{Banados+18}, and hydrogen recombination line masers from the epoch of reionisation \citep{Spaans+Norman97}. Detections will enable a unique investigation of the physical conditions of the intergalactic medium at the end of the epoch of reionisation and the cold gas feeding massive galaxy formation in the early Universe. 

\subsection{The cosmic web and galaxy clusters}
\label{clusters}
\subsubsection{The synchrotron cosmic web}
\label{sec:cosmic_web}
The large-scale structure of the Universe requires the presence of intergalactic shocks, which are in turn expected to accelerate electrons and amplify intergalactic magnetic fields \citep{2004ApJ...617..281K,2005ApJ...631L..21B,2007MNRAS.375...77H,2009MNRAS.393.1073B,2012MNRAS.423.2325A}. These shocks should thus produce faint synchrotron emission, which can act as a tracer of large-scale structure, cosmic filaments and primordial magnetic fields \citep{2004NewAR..48.1119K,2004NewAR..48.1281W,2009MNRAS.392.1008D,2015aska.confE..97V,2015A&A...580A.119V,2017MNRAS.468.4246B}. Detection of this ``synchrotron cosmic web'' can provide a direct image of the large-scale structure of the Universe, act as a laboratory for studying particle acceleration in low-density shocks, lead to a measurement of the magnetic field strength of the intergalactic medium, and provide a direct discriminant on competing models for the origin of cosmic magnetism. It is predicted that the signal from the synchrotron cosmic web should dominate other radio signals on scales of $\sim10'$ to $\sim1^\circ$ at frequencies around 100~MHz \citep{2004ApJ...617..281K,2015aska.confE..97V}, making the MWA a well-suited facility to search for these structures.

\cite{2017MNRAS.467.4914V} have carried out a search for the synchrotron cosmic web with the Phase~I MWA, in which they cross-correlated diffuse radio emission imaged at 180~MHz with large-scale structure traced by infrared galaxy surveys.
They were able to place upper limits on the surface brightness of the synchrotron cosmic web of 0.01\,--\,0.3\,mJy\,arcmin$^{-2}$, which translates to upper limits on the magnetic field strength of 0.03\,--\,1.98\,$\mu$G, assuming equipartition. While these constraints are not yet deep enough to differentiate between different cosmic magnetism models, the limits are comparable to other limits from cluster observations \citep[e.g.,][]{1999A&A...341...29F, 2001MNRAS.320..365B, 2010MNRAS.402....2B} or predictions from MHD simulations \citep[e.g.,][]{2009MNRAS.392.1008D, 2015A&A...580A.119V}.

The depth of the search reported by \cite{2017MNRAS.467.4914V} was limited by confusion, in that large numbers of unresolved extragalactic radio sources have not been subtracted from the data, and may mimic diffuse radio emission that traces large-scale structure. As for many other continuum science programs, the improved angular resolution and reduced confusion levels offered by MWA Phase~II will enable much deeper searches for the synchrotron cosmic web, whether by direct imaging \citep[e.g.,][]{2007ApJ...659..267K}, statistical cross-correlations \citep{2017MNRAS.467.4914V,2017MNRAS.468.4246B} or also in polarimetry \citep{2009AJ....137..145R,2009IAUS..259..669B}.

\subsubsection{Diffuse emission in galaxy clusters}

In addition to the potential detection of the large-scale, diffuse emission predicted by the cosmic web, the Phase~I MWA has proven itself to be a powerful instrument for the detection and study of diffuse, low-surface brightness emission in galaxy clusters. 

Diffuse emission in clusters manifests in a variety of forms including central radio halos, which are believed to be the result of turbulence in the cluster core; mini-halos powered by central AGN; peripheral radio relics which result from strong shocks generated in cluster mergers; and radio phoenices which trace the passage of these shocks across the lobes of AGN (see \citet{Kempner04} for the taxonomy of cluster sources and \citet{BJ14} for a review of the physics of emission in galaxy clusters). 
\change{Recently, \citet{Govoni:2019} observed diffuse emission at \SI{140}{\mega \hertz} connecting the clusters Abell~0399 and Abell~0401 with LOFAR.}

Although on different physical scales from hundreds of kpc (mini-halos and phoenices) to Mpc-scale (halos and relics), one defining characteristic of these sources is their steep radio spectral indices. In the case of halos the average spectral index is $-1.3$, with some examples with spectral indices steeper than $-2$ being detected in MWA data \citep{Duchesne18}. Such emission characteristics make the detection of diffuse cluster emission considerably easier at low-frequencies and the large physical scales make detection easier with instruments sensitive to angular scales of the order of arcminutes. The MWA is thus ideal, and a large number of new and existing sources of diffuse emission in clusters and groups have already been detected and studied with the Phase~I array \citep[][Johnston-Hollitt et al. submitted]{Hindson14,Gerrit17,Duchesne18}. 

\begin{figure}
\includegraphics[width=\linewidth]{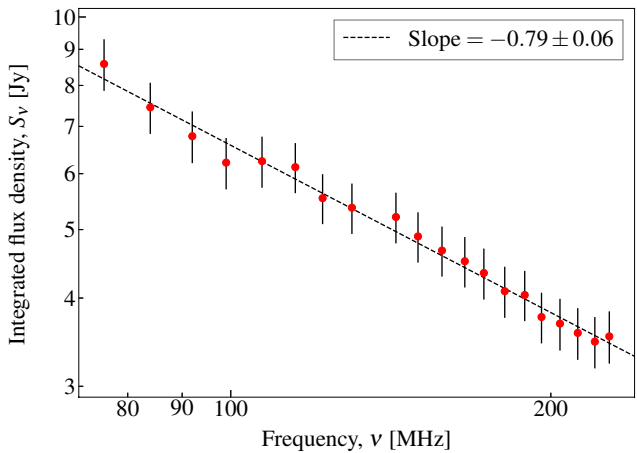}
\caption{MWA data for the complex cluster source NGC 741, showing the unprecedented ability of the MWA to determine spectral indicies across the MWA band. Further information and analysis of these data are available in \cite{Gerrit17}.}
\label{NGC741}
\end{figure}

A particular strength of the MWA is the ability to perform detailed spectral studies of these sources across the MWA band. Figure \ref{NGC741} shows the in-band spectral measurements and resultant fit for the source complex extended cluster source NGC 741 taken from the MWA's GLEAM survey \citep{GLEAM}. This plot demonstrates the extremely fine spectral resolution of the MWA as compared to existing arrays (for a full analysis of these data see \citealp{Gerrit17}). Having such high fidelity spectral information has been invaluable in constraining the physical processes at work to generate the emission, allowing separation of sources that would otherwise be morphologically indistinct into classes based on the underlying acceleration mechanisms and age of the electron population \citep[][Johnston-Hollitt et al. submitted]{Gerrit17}. 

\subsection{AGN and star-forming galaxies}

\change{The population of radio continuum sources is dominated by radio-loud AGN at bright fluxes with an increasing fraction of star forming galaxies (SFGs) as radio surveys reach lower sensitivities \citep{Seymour:2004}.
Consequently the first extra-galactic data release of GLEAM \citep{hw17} contains mostly radio-loud AGN with only a few local SFGs detected.
The wide area of GLEAM and the broad-band radio spectral energy distributions (SEDs) have allowed unique classifications of radio sources by their SEDs \citep{Harvey:2018}.
Detailed studies of radio-loud AGN \citep{Herzog:2016, Callingham+17} and local powerful SFGs \citep{kapinska17, Galvin:2018} are therefore possible. 
However, the depth of GLEAM was limited by the confusion limit from the \SI{3}{\kilo \meter} baselines.
With MWA Phase~II the confusion limit drops by over an order of magnitude allowing far deeper surveys to be conducted.}

\change{\subsubsection{AGN evolution}}

\change{The long baselines have allowed us to conduct new broad-frequency continuum surveys: the eXtended GLEAM survey (GLEAM-X) and the MWA Interestingly Deep Astrophysical (MIDAS) survey.
GLEAM-X is a repeat of the all-sky GLEAM, but with longer total integration in addition to the higher resolution (reaching more than $\sim 5\times$ the depth of GLEAM).
MIDAS is going at least twice as deep as GLEAM-X, targeting six well-studied extra-galactic fields: the five Galaxy And Mass Assembly (GAMA) survey fields \citep{driver+09} and the Spitzer South Pole Telescope Deep Field \citep{Ashby:2014}.
The GAMA survey fields}
cover 250 deg$^{2}$ with exquisite photometry from UV to the far-IR. These fields also include spectroscopic redshifts, group catalogues and derived data products such as stellar mass and star formation rate. The fusion of these data sets will allow a plethora of science including determining the luminosity function (e.g. G\"urkan et al. in prep), and the nature and evolution of the low-frequency selected radio population.

In concert with deep ongoing surveys from ASKAP \citep{norris+11}, uGMRT \citep{Swarup+91} and the Australia Telescope Compact Array (ATCA; GAMA Legacy ATCA Southern Survey$-$GLASS) MWA observations will yield broad-band SEDs for tens of thousands of sources in the GAMA fields.
\change{A large fraction of these will have spectroscopic redshifts and derived host galaxies properties. This unique data set will provide estimates of jet powers for these radio-loud AGN which can be compared to other processes such as accretion rate and star formation \citep[e.g.][]{gg18}.}



\begin{figure}
    \centering
    \includegraphics[width=\columnwidth]{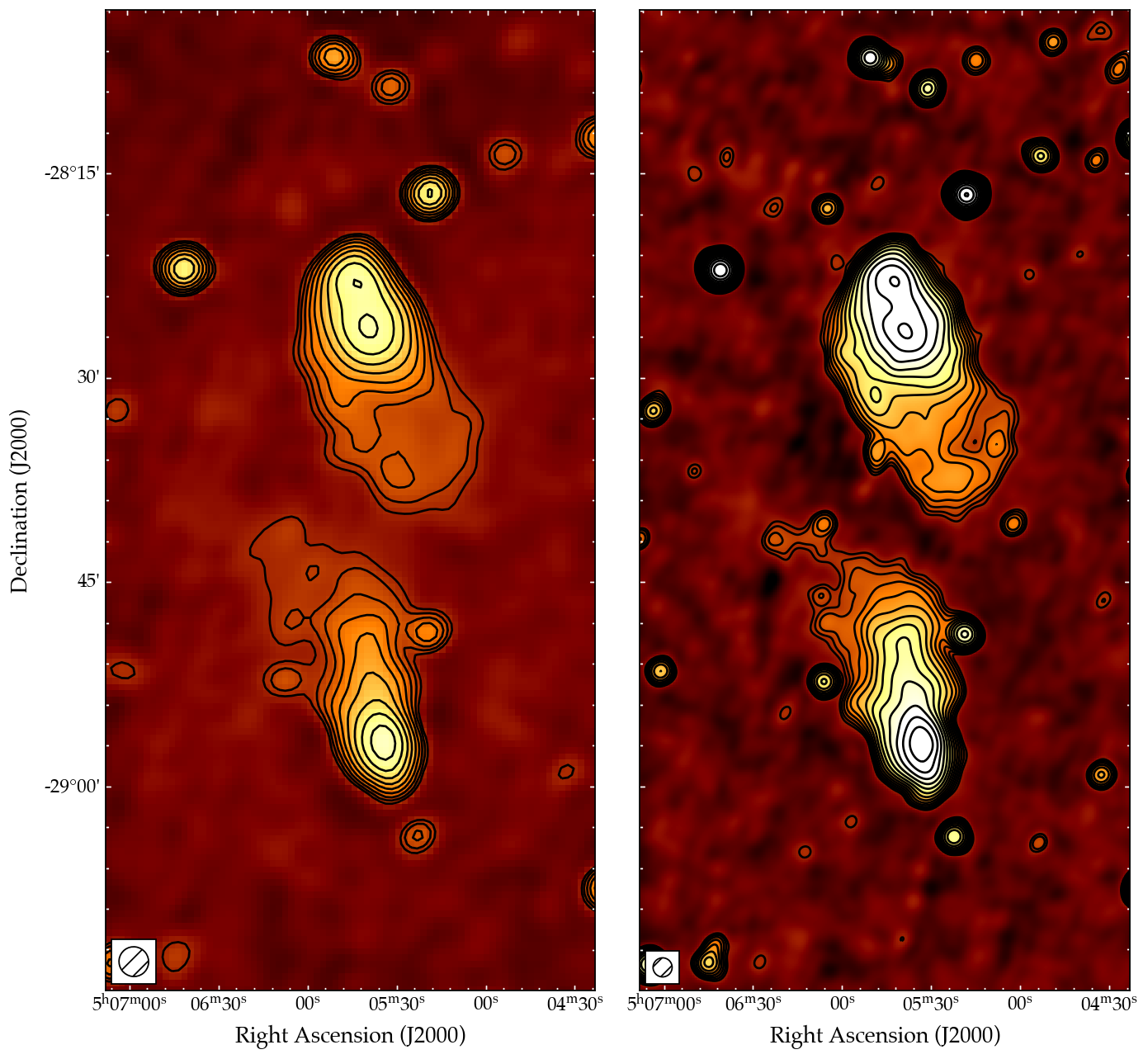}
    \caption{\change{Images of ESO~422$-$G028 from Phase~I (\emph{left}, from GLEAM; \cite{hw17}) and Phase~II (\emph{right}, from GLEAM-X, Hurley-Walker et al., in prep). Colour-scale and contours both denote the surface brightness at \SI{200}{\mega \hertz}. Colour scale ranges from $-3\sigma$ to $100\sigma$ on an arcsinh stretch to emphasise faint diffuse emission. Contours start at $3\sigma$ and scale by a factor $\sqrt{2}$, where $\sigma=14.9$ mJy beam$^{-1}$ ($2.0$ mJy beam$^{-1}$) in the Phase~I (Phase~II) image. The beam size is shown by the hatched ellipse in the lower-left corner.}}
    \label{fig:chris_figure}
\end{figure}

\change{The improvement in sensitivity is also beneficial to the study of Giant Radio Galaxies (GRG), which are AGN with emission that extends over $\sim$ \SI{700}{\kilo \parsec}.
Fine details of this extended emission can now be resolved with the addition of the long baselines, allowing for studies of spectral properties such as the spectral ageing of the lobes, enabling detailed modelling of source environments.
As an example Figure~\ref{fig:chris_figure} compares images of the giant radio galaxy ESO\,422$-$G028 using Phase~I and Phase~II data.
The Phase~II data were imaged using \texttt{robust} $=+1.0$ to improve sensitivity to ultra-diffuse large-scale emission from this GRG.
The improvement in sensitivity and resolution offered by Phase~II of the MWA is illustrated by the ultra-faint diffuse emission from the tails of ESO~422  recovered at much greater significance compared to Phase~I.
With the factor $\sim2-3$ improvement in resolution offered by Phase~II, such data can be used in conjunction with higher-frequency data from other cutting edge radio telescopes (e.g. uGMRT, ASKAP, and the ATCA; Riseley et al., in prep) to probe the spectral properties of extended radio galaxies at much improved angular resolution compared to Phase~I.}

\subsubsection{Powerful AGN}
\change{
The GLEAM 4-Jy Sample \citep{White:2018}, selected at \SI{151}{\mega \hertz}, contains 1,863 bright radio sources} and is dominated by AGN.
This includes 77 radio galaxies that are so extended that they are resolved into multiple GLEAM components by the MWA Phase~I beam. However, approximately one fifth of the sample suffers from confusion \change{which will be alleviated by higher resolution allowing them to be more easily classified and cross-matched.
The 4-Jy sample will be a benchmark for the bright radio galaxy population, like 3CRR \citep{Laing:2003}, but is an order of magnitude larger. }

\change{GLEAM has also allowed us to search for powerful very high-redshift sources through a number of techniques including ultra-steep spectrum, curved SED, and compactness (from IPS, see Section~\ref{sec:IPS}).
This on-going work to obtain their redshifts is proving fruitful with follow-up on \SI{8}{\meter} class telescopes and with the Atacama Large Millimetre Array \citep[e.g.][]{Seymour:2019}.} 

\change{
\subsection{Star-forming Galaxies}
While SFGs are rare in GLEAM, one of the dramatic results to come out of their study was the fact that their SEDs are more complex than had generally been appreciated with flattening and even turn-overs at low frequency as well as kinks at high frequency \citep{kapinska17, Galvin:2018}.
These features are thought to be due to multiple star-forming components, and free-free absorption.
Understanding how this complexity is related to the under-lying star formation is important as using deep radio surveys to trace the star formation history of the Universe is key science goal of the SKA \citep{Braun:2015}.}

\subsubsection{Compact Steep Spectrum and Gigahertz-peaked Spectrum Radio Sources}
Compact steep spectrum (CSS) and gigahertz-peaked spectrum (GPS) radio sources are a class of compact radio-loud AGN that are thought to be the young precursors to large-scale radio galaxies \citep{odea98}. The lower radio luminosity ($<$ 10$^{25}$ W/Hz at 5 GHz) sample of GPS and CSS sources have been suggested to be dominated by the objects strongly affected during their evolution by an interaction with the interstellar medium or instability in the accretion disc \citep{Czerny+09,Bicknell+18}. Such sources could be the short-lived precursors needed to account for the overabundance of GPS and CSS sources relative to the large scale radio galaxies \citep{kb+10}. However, none of the low-luminosity peaked-spectrum samples to date have been large enough, or devoid of selection biases, to justify this conclusion. The increased sensitivity of MWA Phase~II will be able to probe a significantly fainter population of peaked-spectrum sources than previously possible \citep{Callingham+17}. 
\change{While other telescopes such as LOFAR have the required sensitivity to access the low-luminosity peaked-spectrum population, LOFAR currently lacks the wide fractional bandwidth needed to identify peaked-spectrum candidates nontrivially when solely using LOFAR data.
Therefore, the large fractional bandwidth of the MWA, combined with the high sensitivity of Phase~II, represents a significant advantage for this science.}

\subsubsection{Dying radio galaxies}

As with the diffuse emission in galaxy clusters, the MWA has a niche capability in the detection of diffuse, low surface brightness emission from the lobes of old, dead, or dying radio galaxies. One such early example was the detection of previously unknown giant radio lobes associated with the lenticular galaxy NGC 1534 \citep{HW15, Duchesne:2018}. Work in this field across a range of low frequency telescopes has seen the detection of several more such sources, suggesting a significant and previously unknown population of sources (e.~g.~ \citealp{deGasperin14,Brienza16,mahatma+18}). However, as with the diffuse emission in galaxy clusters, the volume over which remnant or dying radio galaxies can be detected was limited with the Phase~I array to the very local Universe. If we assume a typical size for such sources of 700 kpc, in order to resolve the lobes of such galaxies, the Phase~I MWA would have been limited to sources with redshifts less than 0.17. Even detecting such sources in the nearby Universe, the Phase~I MWA had insufficient resolution to perform spectral imaging across the lobes of the radio sources, which is vital to understanding the physics of such sources. As a result while the Phase~I MWA was able to detect some of this class of object, no detailed imaging could be undertaken. 
The detection limits with the Phase~II array can be pushed out to redshifts of z $\sim$ 0.42, which considerably expands the volume in which such sources can be found. Furthermore, there will be sufficient resolution to perform detailed statistical spectral studies over the radio lobes of nearby sources, which is vital to understand the physics of such systems.


\subsection{Polarimetry}

The detection and characterisation of polarised sources in the low frequency sky is a challenging endeavour due to the Faraday depolarisation of radio sources with increasing wavelength. This is further compounded in the case of low resolution instruments where beam depolarisation can occur. Nevertheless, polarimetry campaigns are on-going across a number of low-frequency instruments including \change{LOFAR}, PAPER, and the upgraded GMRT \citep{2014A&A...568A.101J,Kohn2016,VanEck2018, 2018Galax...6..126O}. 
The MWA's wide field of view and broad frequency coverage make it a highly efficient facility for the detection of polarized radio sources, highly precise determinations of their Faraday rotation measures (RMs), and the study of Faraday complexity.

Exploration of the polarised sky with the MWA commenced with the 32-tile prototype array \citep{2013ApJ...771..105B} and has continued for both compact extragalactic \citep{Lenc+17, 2018PASA...35...43R} and diffuse, wide-field Galactic emission with the Phase~I array \citep{Lenc2016}. 
The total number of discrete polarized point sources detected with the Phase~I MWA was expected to be low due to limited sensitivity and angular resolution ($\sim250$ linearly polarized sources estimated across the visible Southern sky; \citealt{Lenc+17}). 
Re-analysis of the Phase~I GLEAM survey data as described by \citet{2018PASA...35...43R} has proven to be an effective method of identifying and studying such sources, and has now yielded about a factor of two higher source density of extragalactic polarized sources than projected, in addition to tens of pulsars detected in polarization after targeted inspection (Riseley et al., in prep.).
All of the extragalactic polarized sources are AGN.
Many are detected in polarization from the hotspots of Faranoff-Riley type II (FRII) sources, and some are interesting multi-component cases such as PKS J0636$-$2036, which is the first source for which a broadband polarimetric study (wavelength-squared coverage from 1.7 to 16 m$^2$) has been undertaken using combined MWA and Australia Telescope Compact Array imaging \citep{O'Sullivan18}. 
Detailed study of the Faraday complexity of polarized sources detected with the MWA is possible by combining data from ASKAP's forthcoming Polarisation Sky Survey of the Universe's Magnetism \citep[POSSUM;][]{Gaensler+10} survey and the ATCA QUOCKA survey (Heald et al., in prep.) which spans the \SIrange{1}{8.5}{\giga\hertz} frequency range.

The Phase~II MWA is expected to detect vastly more compact polarised sources through both the reduction in beam depolarisation provided by the longer baselines, and the improvement in sensitivity, particularly for imaging modes
where the longest baselines are not significantly down-weighted \citep{Lenc+17}.

In addition, the improved resolution of the Phase~II MWA will allow a more detailed exploration of the vast swaths of diffuse, linearly polarised Galactic emission detectable by the MWA \citep{Lenc2016}. Study of the diffuse Galactic polarized emission, together with Faraday rotation measures of extragalactic sources and polarized pulsars, allows holistic modeling of the three-dimensional Milky Way magnetic field. For example, RMs of pulsars embedded within the Galactic halo provide constraints on the vertical change in magnetic field strength \citep{2019MNRAS.484.3646S}.

Finally, consideration of circular polarisation measurements with the Phase~II MWA will detect flare stars and has the potential to make the first low-frequency detection of exoplanets. In the case of flare stars, the Phase~I MWA has already verified the potential for faint flares to be detected in polarised images. \citet{Lynch2017b} detected four flares from UV Ceti with the Phase~I array, each with flux densities over two orders of magnitude fainter than previously reported. Due to confusion in the total intensity imaging, these faint flares were only detectable in the polarised images, demonstrating the power of low-frequency arrays to perform such searches in circular polarisation. The improved characteristics of the Phase~II MWA should thus allow for more such detections and an assessment of their temporal characteristics. Furthermore, although previous attempts to directly detect exoplanet emission with the MWA via circular polarisation have been unsuccessful, setting only upper limits \citep{Murphy15,Lynch2017a}, the experiment should be repeated with the Phase~II array which is expected to be an order of magnitude more sensitive as discussed in Section \ref{sec:cosmic_web}. We further discuss MWA exoplanet searches in the context of transient science in Section~\ref{sec:exo}.

\subsection{Complementarity with other Surveys} 
The MWA upgrade is taking place in the context of a revolutionary period in radio astronomy, with several new large radio surveys taking place on major new or dramatically upgraded telescopes around the world, resulting in an increase by a factor of $\sim$50 in the number of known radio sources \citep{norris17}. 
\change{Of particular relevance to MWA surveys is the northern-hemisphere LOFAR Two-Metre Sky Survey \citep[LoTSS][]{Shimwell+17}, with similar frequency band (\SIrange{120}{168}{\mega \hertz}) and higher angular resolution ($\sim$\SI{5}{\arcsecond})}
The SKA precursor continuum surveys MIGHTEE \citep{jarvis+16} on MeerKAT \citep{Jonas9}, and EMU \citep{norris+11} and POSSUM \citep{Gaensler+10} on ASKAP \citep{Johnston+08} will provide higher-frequency data for MWA sources, and the VLASS survey \citep[][; Lacy et al., in preparation]{Murphy+15} will provide even higher frequency and high-resolution coverage over the northern half of the MWA coverage. 
Also important is the major upgrade of the GMRT \citep{uGMRT}, which is likely to generate surveys of large parts of the sky from 150 to 1400 MHz.
For total intensity studies of extragalactic source SEDs, the EMU survey on ASKAP is particularly important as it will produce a catalogue of high-frequency counterparts for all sources detected by the MWA. ASKAP can measure the SED from \SIrange{700}{1800}{\mega\hertz}.
Because the ASKAP configuration includes many short baselines, the data can be tapered in the {\em uv} plane to match MWA resolution with only a small loss of sensitivity. Thus, images with matched resolution can be generated at 150-300 MHz with MWA, and 700 -- 1800 MHz with ASKAP, to produce reliable SEDs from 150 to 1800 MHz with only a relatively small gap (300 -- 700 MHz) in frequency coverage.
These broad-band SEDs will obviously be important for studies of clusters (see Section \ref{clusters}), but will be particularly important for studies of radio AGN. Recent studies \citep[e.g.][]{Callingham+17} have shown that a significant fraction of radio sources have curved or peaked SEDs accessible by the joint MWA/ASKAP coverage, and modelling of this (Shabala et al., in preparation) is likely to lead to a good understanding of their evolutionary stage, and perhaps even redshifts. 

This wide frequency range will also be important for polarisation studies. Combining the polarisation data from MWA \citep{Lenc+17} with that from the POSSUM project on ASKAP and the ATCA QUOCKA survey will be ideal for studying Faraday synthesis.
A full understanding of the magnetic field properties of polarized sources detected with the MWA will be provided by detailed modeling of the combined polarizd SEDs \citep[see e.g.,][]{2016ApJ...825...59A, 2018Galax...6..126O}. Here, most of the coverage in the relevant $\lambda^2$ domain comes from the MWA.

\section{TIME DOMAIN ASTROPHYSICS}
\label{sec:timedomain}
Many exciting astrophysics topics of the day lie in the time domain including the deluge of characterized exoplanets, the intriguing nature of FRBs, the first detections of gravitational waves and their electromagnetic counterparts, and indeed most multi-messenger astrophysics.
The MWA's recently deployed rapid-response triggering system and new VCS buffer mode have greatly improved the telescope's ability to react to alerts from other observatories\change{. T}he high resolution of the extended configuration will enable better localization constraints\change{, while the higher correlator data rate enables transient followup at high time resolution and exploration of wider range of dispersion measures of prompt radio signals \citep[e.g.,][]{msok_mwa_askap}}.
On the other hand, the already wide field of view, new compact configuration, continued developement of tied-array beamforming \change{\citep{Ord:2019, Xue:2019}}, and the new BL back-end make the MWA an extremely powerful instrument for blind transient searches. Here we discuss the new science afforded by these improvements.

\subsection{Gamma-Ray Bursts}
\label{sec:grb}

\def\etal{et al}	
\def\swift{\textit{Swift}}
\def\fermi{\textit{Fermi}}

%
%
Gamma-ray bursts (GRBs) are one of the most energetic phenomena in the Universe, hallmarked by their bright flash of gamma-ray emission. GRBs are detected by dedicated satellites such as the \swift\ Burst Alert Telescope \citep[BAT;][]{barthelmy05} and the \fermi\ Gamma-ray Burst Monitor \citep[GBM;][]{meegan09}, which transmit alerts in near-real-time that allow for rapid multi-wavelength follow-up. 
Short-duration gamma-ray bursts \citep[SGRBs, which last $<2$\,s;][]{kouveliotou93} are of particular interest due to the recent, near-simultaneous detection of the gravitational wave (GW) event GW170817 and GRB 170817A \citep[the latter of which was detected by \textit{Fermi};][]{2017ApJ...848L..13A}, which strongly supports the proposed link between binary neutron star (BNS) mergers and SGRBs \citep{eichler89}.
Detailed multi-wavelength studies of SGRBs can therefore shed light on GW afterglows by providing a template of their brightness and timing properties that will inform the follow-up of future GW events by wide-field instruments.

GRBs afterglows,
which are generated by the interactions of their relativistic jets with the surrounding medium, are also known sources of radio emission in the GHz range \citep{piran99}. 
However, it is unlikely that such afterglows will be detected by the MWA as the synchrotron emission produced at MHz frequencies only peaks after hundreds to thousands of days and can be quite faint. 
Instead, we examine the fact that BNS mergers, and therefore SGRBs, may produce prompt, coherent emission \citep{2013PASJ...65L..12T,2014AA...562A.137F,2014ApJ...780L..21Z}, and may therefore be responsible for at least some (non-repeating) FRBs
\citep{lorimer07,2013Sci...341...53T}.
In fact, the recent detection of a candidate gamma-ray transient temporally and spatially coincident with an FRB provides a further suggestion for such an association \citep{delaunay16}. 
Such prompt emission may be related to the initial merger or the creation, lifespan, and collapse of a short-lived ($\lesssim10^{4}$\,s), supramassive and highly magnetised NS, commonly referred to as a magnetar \citep{usov92,rowlinson13,ravi14}. 
The initial interaction of the gamma-ray jets with the surrounding medium may also produce FRB-like emission \citep{2000AA...364..655U}. Many of these theories may also hold for long-duration GRBs \citep[which are the collapse of massive stars into black holes and last $>2$\,s;][]{kouveliotou93}, such as prompt emission from jet interactions or the formation of magnetars \citep{bernardini12}. However, it is unclear if any pulsed coherent emission would be detected due to their higher density environments \citep{lyubarsky08}. 

A summary of prompt radio emission models from BNS mergers can be
found in \citet{2016MNRAS.459..121C}, with additional models proposed
later \citep{2016ApJ...822L...7W,2016MNRAS.461.4435M,2018arXiv180804822P}. 
The detection of emission associated with SGRBs would distinguish between different binary merger models and in-turn constrain the equation-of-state of nuclear matter \citep{lasky14}. 

As pulsed radio emission arrives at later times with decreasing frequency due to dispersion by the intergalactic medium, low-frequency ($<300\,$MHz) radio telescopes like the MWA are ideal for probing these prompt signatures, provided they can be on source within seconds to minutes following the outburst.

The rapid-response system described in Section~\ref{sec:rapid} is capable of receiving external transient alerts transmitted via the VOEvent standard and automatically repointing the telescope, beginning observations of the transient event within \SI{14}{\second} of receiving the alert \citep{fast_response}.
%
With such a quick repointing response, combined with the expected dispersion delay of $10-100$\,s at 185\,MHz within the redshift range of $0.04 \leq z \leq 0.7$ (spanning the aLIGO/Virgo expected sensitivity limit of 200\,Mpc and the average red-shift of \swift\ SGRBs; \change{\citealt{Rowlinson:2019}}), the MWA is capable of being on target in-time to probe FRB-like emission associated with GRBs. Such prompt signals are also expected to smear in time at the above redshift range, taking between $5-30$\,s to cross the MWA's 30\,MHz bandwidth (for $170-200$\,MHz). An additional advantage of the MWA is its wide field of view, which makes it one of the few telescopes capable of covering the positional uncertainties ($\sim10$\,deg) of \fermi\ GRBs \citep{2016PASA...33...50K}. 

The original rapid-response system triggered on SGRB\,150424A within 23\,s following the \swift\ detection, placing some of the most stringent limits on pulsed radio emission from GRBs at MHz frequencies with MWA Phase~I \citep[3\,Jy on 4\,s timescales;][]{2015ApJ...814L..25K}. Many of the prompt emission models predict single radio pulses that reach Jy levels from GRBs at $z=0.7$ \citep[e.g.][]{2014ApJ...780L..21Z}, and are therefore likely detectable with any MWA configuration, none of which are confusion limited on second timescales. However, it is unclear what sort of flux levels to expect from prolonged pulsar-like emission produced by a magnetar formed during the BNS merger, which is dependent on the efficiency ratio of the radio luminosity to its total energy loss rate \citep{2013PASJ...65L..12T}. As the Phase~II extended configuration is not confusion limited in \SI{30}{\minute}, observations following up on SGRB alerts will be far more sensitive than Phase~I, and therefore more capable of detecting such a signature. Both the standard correlator imaging mode and the VCS, applying appropriate de-dispersion measures, will be used to search for these pulsed signals from GRBs.

\subsection{Multi-Messenger Astrophysics}
Astrophysical signals beyond electromagnetic (EM) radiation provide unique windows into the universe. 
The wide field of view, high sensitivity, and rapid triggering response of the MWA afford opportunities to complement observations of gravitational waves, neutrinos, and high energy cosmic rays. 
Here we outline efforts underway to detect low-frequency EM counterparts.

\subsubsection{Gravitational Waves}
\label{sec:gw}

In 2016 the LIGO/Virgo Consortium (LVC) reported the first two detections of gravitational waves \citep{ligo16, ligo16b}.
The LVC sent private alerts to the EM followup community \citep{2016ApJ...826L..13A,2016ApJS..225....8A} to identify coincident
EM transients.
While identifying an
EM counterpart 
would greatly enhance the utility of the GW signal
\citep[e.g.,][]{2009astro2010S.235P,2012ApJ...746...48M, 2014ApJ...795..105S,2016MNRAS.459..121C,2016arXiv160408530B}, it
is not a simple task owing to very large uncertainty regions for the GW events, which can be hundreds of degrees
\citep[e.g.,][]{2014ApJ...789L...5K, 2014ApJ...795..105S}.

As mentioned in Section~\ref{sec:grb}, this hard work bore amazing fruit in 2017,
when LVC detected a neutron star merger GW\,170817 \citep{2017PhRvL.119p1101A}
coincident with a short (underluminous) gamma-ray burst, GRB\,170817A
\citep{2017ApJ...848L..13A} which was subsequently detected across the EM
spectrum \citep[][where MWA 
participated]{2017ApJ...848L..12A}.  
While this one event has generated a wealth of
information, there are still fundamental questions as to the nature of
the multi-wavelength emission, the energetics of the explosion, the
properties of the environment, and the degree of beaming
\citep[e.g.,][]{2017Sci...358.1559K,2018arXiv180609693M,2018Natur.554..207M,2018MNRAS.479..588G,2018MNRAS.478L..18T,2018PhRvL.120x1103L,2018ApJ...863L..18A,2018ApJ...858L..15D}.  Many of these can only be answered by looking at a much larger sample of objects, and 
it must be determined what strategies and types of signals will give the 
best return in the radio domain.  

A number of models for
neutron star-neutron star mergers predict a prompt flash of coherent
radio emission that should accompany (slightly before or after) the GW
signal and be detectable by the MWA (see Section~\ref{sec:grb} for specific examples).
In fact, low frequency telescopes have a number of 
advantages over optical/infrared searches: 
they have fields-of-view of
hundreds to thousands of square degrees; 
the radio
sky is relatively quiet at these frequencies
\citep{2015MNRAS.452.1254K,2015AJ....150..199T,2016MNRAS.456.2321S,2016MNRAS.458.3506R,2016ApJ...832...60P}
with very few transients unassociated with the GW event
\citep[e.g.,][]{2016ApJ...831..190H}; 
and  \change{owing to the new rapid-response mode,} the MWA can respond within seconds
to an external trigger.  
A detection of a prompt coherent flash would
give immediate localization of the GW signal with a very low
false-positive rate, and would help probe the distance to the source
through constraints on the line-of-sight electron density if
dispersion can be measured.  
Unfortunately, 
GW\,170817 was not visible to the MWA when the trigger was announced,
and observations from other sites were not sufficiently sensitive
(\citealt{gcn21975}, using the LWA, was a factor of $\sim 100$ less sensitive than
typical MWA observations discussed below).

Along with the VOEvent triggering mechanism discussed in
Section~\ref{sec:rapid},
an optimal
strategy for MWA followup of prompt emission from GW transients has been implemented \citep{2016PASA...33...50K}.
This is complementary to
approaches taken with other facilities
\citep[e.g.,][]{2015ApJ...812..168Y} where the wide FoV, rapid response, and Southern location of
the MWA give it an advantage \citep{2016MNRAS.459..121C,2015PASA...32...46H}.
\citet{2016PASA...33...50K} used simulated GW events from
\citet{2014ApJ...795..105S}  to compute
the expected fraction of events that the MWA would observe and the
sensitivity to them, given the optimized
pointing strategy.
For optimum conditions, the limiting (10\,$\sigma$) flux density is
$\sim 0.1$\,Jy.  However, given the influence of Galactic synchrotron
emission and the limited collecting area away from zenith, only 5\% of
simulated events are close to that limit; a more typical sensitivity
is 1\,Jy.  For typical distances ($\sim 200\,$Mpc), that corresponds to luminosity
limits of $10^{38-39}\,{\rm erg/s}$.  These luminosities
are squarely in the range of various
predictions for prompt radio emission
\citep{2000AA...364..655U,2010ApSS.330...13P,2013PASJ...65L..12T,2014ApJ...780L..21Z,2014AA...562A.137F},
and should seriously constrain the underlying physics
\citep{2015ApJ...814L..25K} even with a non-detection.
Note that the MWA's rapid (\SIrange{6}{14}{\second}) repointing is crucial to capturing
prompt emission, because dispersive delays of only tens of
seconds are expected at the MWA frequencies.  During the LIGO/Virgo O3 run (started 2019 April) there is an expectation of low-latency alerts with reduced information, which will be ideal for triggering MWA followup. Note that if prompt gamma-ray
emission is detected again the system will trigger on the GRB
itself, since that notice is sent within seconds of the event.

\change{
Recently, \citet{James:2019} proposed  triggering the MWA on negative latency GW alerts (generated from the GWs emitted by the inspiral of the binary components prior to merger).
These alerts have relatively poor positional constraints, and so this observing mode would disable 15 of the 16 dipoles on each tile, resulting in a FoV covering about a quarter of the sky.
This mode is expected to improve the MWA's response time by several seconds, significantly increasing the potential to detect FRB-like prompt emission from mergers.
}


\subsubsection{Neutrinos}
Multimessenger astrophysics at MWA also involves searching for counterparts to astrophysical neutrino sources. 
In addition to archival searches for serendipitous observations of neutrino triggers (Section~\ref{sec:archive}), MWA is  undertaking pointed observations to follow up triggers from neutrino observatories including IceCube \citep{icecube} and ANTARES \citep{antares}. 

Recent multi-wavelength follow-up observations of an IceCube neutrino alert identified the blazar TXS~0506+056 as the first detected high-energy ($E>$\,\SI{e12}{\electronvolt}) astrophysical source \citep{icecubeblazar}. 
However, the mechanism behind the neutrino emission is unknown, with the most optimistic models predicting only 1\% of the observed signal \citep{2018ApJ...864...84K}. 
Further, no more than 27\% of IceCube's high-energy astrophysical flux can be explained by blazar emission \citep{2017ApJ...835...45A}.

EM follow-up of neutrino events to identify the remaining neutrino sources is therefore a priority. Candidate astrophysical sources of the remaining flux of high-energy neutrinos include GRBs, core-collapse SNe, microquasars, or AGN. The recent observation by the ANITA Antarctic balloon experiment of two $\sim$\SI{1}{\exa \electronvolt} neutrino events in apparent contradiction of standard model physics also serves as a reminder that neutrino observations could point the way to as-yet unknown phenomena \citep{2016PhRvL.117g1101G}.

Similar to GWs, the wide FoV\change{, increased resolution and sensitivity, and new rapid-response mode} of the MWA are well suited to react to neutrino alerts. IceCube and ANTARES are expected to generate a total of 6\,--\,8 alerts per annum with a position visible to the MWA. Triggering on these events and performing the subsequent observations will allow for the strongest limits to date on prompt radio emission from neutrino
transients and may aid in localization of these new astrophysical probes.


\subsubsection{Cosmic Rays}
\label{sec:cr}

The origin of cosmic rays is unknown. In the region of the energy spectrum between the `knee' at $10^{15}$--$10^{16}$\,eV, and the `ankle' at $10^{18}$--$10^{19}$\,eV, cosmic rays are thought to transition from a predominantly Galactic to an extragalactic origin.  In this region, experiments aim to measure the spectrum of each primary particle species (p, He, CNO, Fe etc.) \citep{2013APh....47...54A}, with spectral features identifying Galactic accelerators via the magnetic rigidity of the primaries \citep{Peters1961}. 
Using the low-band antennas (LBAs) from its inner stations, LOFAR has established that a dense array of low-frequency radio antennas can accurately reconstruct key cosmic ray properties.

When a cosmic ray impacts the upper atmosphere, it produces an extensive air shower of secondary particles, which in turn emit a sub-microsecond burst of radio waves \citep{2016PhR...620....1H}. The frequency structure and ground pattern of these bursts depend on the properties of the particle cascade and, hence, on the primary cosmic ray itself.

By studying the ground pattern of cosmic ray cascades \citep{2013A&A...560A..98S}, radio data from LOFAR has allowed the height of shower maximum, $X_{\rm max}$, to be reconstructed with unprecedented accuracy \citep{2014PhRvD..90h2003B}. The importance of $X_{\rm max}$ is that it can be statistically related to the cosmic ray composition.

LOFAR's detection threshold of approximately $10^{16}$\,eV lies in the Galactic--extragalactic transition region, and its measurements have indicated a new light-mass component to the flux in the $10^{17}$--$10^{17.5}$\,eV range \citep{2016Natur.531...70B}.

The MWA's bandwidth of $30.72$\,MHz and flatter bandpass will provide a comparable sensitivity to the $30$--$80$\,MHz range used by LOFAR's Cosmic Ray Key Science Project. The rate of cosmic ray detections is also expected to be comparable to that of LOFAR \change{due to the concentrated sensitivity of the compact configuration of Phase~II}. The core region, including hexes, spans an area similar to that of the LOFAR superterp. Beamforming $N_{\rm dip}$ (here, 16) dipoles per tile increases signal-to-noise by $N_{\rm dip}^{0.5}$, and reduces the cosmic ray energy detection threshold $E_{\rm thresh}$ by $N_{\rm dip}^{-0.5}$. Since the cosmic ray rate $R_{\rm CR}$ falls with energy $E_{\rm CR}$ approximately as $\frac{dR_{\rm CR}}{dE_{\rm CR}} \sim E_{\rm CR}^{-3}$ in this range, the integrated \change{rate} varies as $E_{\rm thresh}^{-2}$, i.e.\ with $N_{\rm dip}$. This exactly compensates the $N_{\rm dip}$-fold loss of solid angle from beamforming, so that the rate is constant. This will therefore allow the MWA to probe lower-energy regions of the cosmic ray spectrum with similar statistics to LOFAR, and test predictions of a rigidity-dependent spectrum.

\begin{figure}
{
\includegraphics[width=\columnwidth, trim=110 0 50 0, clip]{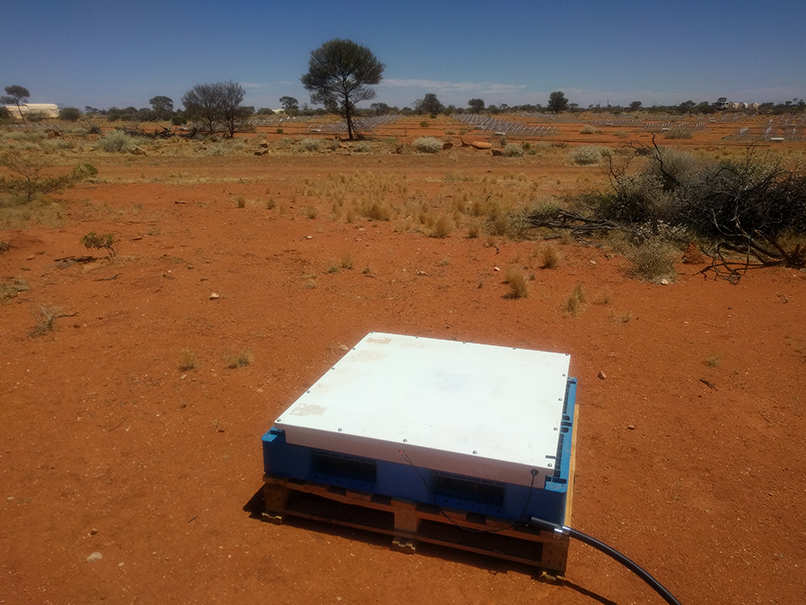}
}
\caption{\change{The prototype SKAPA cosmic ray detector (white box) deployed near the MWA core at the MRO. It is raised off the ground with palettes to avoid surface water. Power and data return are enclosed via the black cable housing. The detector has been tested to ensure compliance with MRO RFI requirements, and has been detecting cosmic ray muons since its deployment in October 2018.  \emph{Photo credit}: Justin Bray.}
\label{fig:crd}
}
\end{figure}

Identification of cosmic ray signals at the MWA site could be performed using either on-site particle detectors, or triggered on radio-only data. 
\change{Particle detectors are being developed for SKA1-Low, and tested with the MWA (Figure~\ref{fig:crd}). 
An array of eight  detectors is planned to trigger MWA radio observations of high-energy cosmic ray events.}
While \change{radio-only triggering} has proven infeasible for most modern experiments, \change{it} has recently been tested successfully at the Owens Valley Radio Observatory Long Wavelength Array \citep{Monroe:Thesis}. The feasibility of both \change{techniques} are currently under investigation, in particular the resynthesis of time-domain data from $24$ coarse channels provided by the VCS.

While cosmic ray emission below $80$\,MHz is well-understood, the frequency range above $80$\,MHz is relatively unexplored, with the only observations being from experiments measuring the radiation pattern at essentially a single point (e.g.,\ the Antarctic Impulse Transient Antenna, ANITA \citealt{2010PhRvL.105o1101H}; Cosmic-Ray Observation via Microwave Emission, CROME \citealt{2014PhRvL.113v1101S}). Simulations predict that $110$--$190$\,MHz is the most sensitive frequency range at which to observe cosmic rays \citep{2018EPJC...78..111B}. This opens the possibility of observing radio emission from the flux of PeV ($10^{15}$\,eV) gamma-rays from the Galactic Centre source detected by the High Energy Stereoscopic System, H.E.S.S.\ \citep{2016Natur.531..476H}. Furthermore, Corsika-based Radio Emission from Air Showers (CoREAS) simulations suggest that the ground pattern in Stokes V (circular polarisation) changes shape at higher frequencies \citep{2013AIPC.1535..128H}. This arises from interference along the cascade axis, and points to a method of resolving air shower structure --- and hence composition --- in more detail than can be provided by $X_{\rm max}$ alone.


\subsection{Fast Radio Bursts}
\label{sec:frb}

FRBs are one of the most exciting astrophysical transient phenomena discovered in recent years \citep[just over a decade ago, e.g. ][]{lorimer07, 2013Sci...341...53T}. The origin of these extragalactic events remains unknown.
FRBs are observed as millisecond-timescale bursts of bright radio emission with high dispersion measures (DM) significantly exceeding DMs of objects in the Milky Way.
Until very recently they have only been observed at frequencies above 700\,MHz \citep{Masuietal2015} with the highest frequency detections at 8\,GHz \citep{2018arXiv180404101G}.
Major efforts have been made by several groups to detect FRBs at low frequencies (e.g.  \citealt{2014A&A...570A..60C}, \citealt{2015MNRAS.452.1254K}, \citealt{2017ApJ...844..161A} and \citealt{2017ApJ...844..140C}). 
Recently the Canadian Hydrogen Intensity Mapping Experiment \citep{2018arXiv180311235T} successfully detected FRBs down to \SI{400}{\mega\hertz} \citep{CHIMEFRB:2019}, and provided evidence that they can be detected close to the upper end of the MWA frequency range (around \SI{300}{\mega\hertz}). 
To date, no FRBs have been detected at frequencies below 400\,MHz.
Moreover, no FRB has been detected simultaneously at high ($>1$\,GHz) and low ($<1$\,GHz) frequencies.

The MWA has been in the forefront of low-frequency FRB searches from its beginning \citep{2015AJ....150..199T,2016MNRAS.458.3506R,2016Natur.530..453K}.
The recent work by \citet{msok_mwa_askap} took advantage of the synergy and geographical co-location of the MWA and Australian Square Kilometre Array Pathfinder (ASKAP) radio-telescopes at the 
MRO to shadow (co-track) the pointing positions of the ASKAP antennas performing the Commensal Realtime ASKAP Fast Transients (CRAFT) survey at 1.4\,GHz \citep{Macquartetal2010}.
The MWA data were collected in the correlated mode at 10-kHz frequency and 0.5-second temporal resolutions and
1.28-MHz / 0.5-second images were formed, de-dedispersed and searched for low-frequency ($170-200$\,MHz) counterparts around the ASKAP error regions. 
This shadowing campaign resulted in simultaneous MWA observations of seven extremely bright ASKAP FRBs. 
No low-frequency counterparts were detected, but three of these events provided the tightest direct constrains on the broadband spectral indexes of FRBs. 

The extended MWA, in conjunction with the ASKAP real-time FRB pipeline, will open a unique opportunity for the first ever low-frequency and broadband detections of FRBs, 
bringing about a significant breakthrough that would trigger the review of FRB models.
Therefore, shadowing of the CRAFT survey will also be continued at the  highest MWA frequencies. 
In addition, the recently implemented automatic triggering system (\autoref{sec:rapid}) will be used.
In this mode, FRBs identified by ASKAP in real-time will trigger an automatic re-pointing of the MWA to the FRB position within $16$\,seconds\footnote{A \SI{16}{\second} dispersion time delay between \SI{1.4}{\giga\hertz} and \SI{200}{\mega\hertz} corresponds to DM$\approx 160$. Almost all FRBs discovered to date have DM$>160$.}, recording voltages at high frequency and temporal resolution using the VCS. 
Furthermore, the Voltage Capture Buffer mode will enable recording of high-resolution voltages up to \SI{150}{\second} before the alert arrival.
These new developments enable acquisition of data with about an order of magnitude higher sensitivity than the previous MWA searches. 
Simultaneous broadband detections (\SI{1.4}{\giga\hertz} and below \SI{300}{\mega\hertz}) would constrain FRB energetics and help answer multiple remaining questions about the FRB progenitors and signal propagation.
Even non-detections will result in much tighter upper limits enabling characterisation of broadband properties of FRBs, establishing the presence of low-frequency cut-off and 
consequently constraining FRB energetics, which will narrow down a number of feasible FRB models.


\subsection{Exoplanets and Other Polarized Transients}
\label{sec:exo}

Magnetised exoplanets are expected to produce detectable, low-frequency radio emission via the electron cyclotron maser instability \citep{winglee86, zarka01}. The radio emission from magnetised exoplanets is expected to be brighter than the radio emission of most host stars \citep{griessmeier05}, and radio observations could provide another way to detect exoplanets directly. Further, radio detections provide the most promising method for a direct measurement of the planet's magnetic field strength. Such a measurement would provide insight into the interior composition of these planets. The variability of detected emission in time and frequency will also provide constraints on the exoplanet's rotational period, orbital period, and the geometry of the magnetic field with respect to the exoplanet's rotation axis \citep{hess11}.

From observations of the magnetised Solar System planets, electron cyclotron maser emission is observed to be highly circularly or elliptically polarised, beamed, and variable on time-scales ranging from seconds to days \citep{wu79, treumann06}. Using scaling relations based empirically on the Solar System planets, several authors have determined that the exoplanets most likely to be detected by current radio telescopes are rare, and orbiting in extreme environments. These systems include hot Jupiters orbiting still forming stars \citep{lazio04, stevens05, griessmeier05, vidotto10}, Jupiter-like planets orbiting giant type stars \citep{fujii16}, and orbiting stars that produce copious amounts of X-ray and UV emission \citep{nichols11, nichols12}. 

Despite the large number of low-frequency searches for radio bright exoplanets, there have yet to be any unambiguous detections \citep{bastian00, lazio04, george07, smith09, lazio10, stroe12, lecavelier13, hallinan13, sirothia14, Murphy15, Lynch2017a, ogorman18, lenc18, lynch18}.  Taking advantage of the expected high circular polarisation of the planetary emission and low confusion noise in Stokes V, MWA Phase~I efforts focused on searches in polarised emission. \citet{Lynch2017a} did a blind search for planetary emission within MWA observations of Upper Scorpius, a young star forming region, and placed the first low-frequency limits on radio emission from planets orbiting still forming stars. \citet{Murphy15} and \citet{lynch18} used MWA all-sky surveys to place limits on low-frequency emission from known exoplanets that were estimated to produce the brightest radio emission. \citet{lenc18} expanded these searches, placing low-frequency limits on all known exoplanets within the sky-coverage of their all-sky circular polarisation survey (see below). 

The expected order-of-magnitude increase in the sensitivity of the MWA Phase~II will move searches for radio bright exoplanets into a sensitivity regime where many more exoplanets are predicted to produce observable levels of radio emission. This is especially important in total intensity as a detection in circular polarisation alone does not help to constrain physical models of the planet's magnetic field structures -- what is needed is the polarisation fraction and any associated variability in the amount of polarised emission \citep{hess11}. The greater number of potentially observable exoplanets will increase the likelihood of making detections, given the expected beaming of the radio emission, and will help to place meaningful constraints on current planetary radio emission scaling relations. 

More  generally, searching for circularly polarized sources is a
promising avenue for a wide range of targets.  Many of the sources
that vary strongly at low frequencies do so because of coherent
processes that are inherently polarized (e.g., pulsars, low-mass
stars, exoplanets, etc.; \citealt{Lynch2017a,lenc18}).  However, the
polarized sky is largely empty of confusing sources \citep{lenc18}, which improves
sensitivity and greatly simplifies source association.  Doing
large-scale repeated surveys for polarized transients is therefore a
promising method of detecting rare, interesting objects but is only
possible with an instrument with a large survey speed such as the
Phase~II MWA.

\subsection{X-ray binaries}
\label{sec:xrb}
X-ray binaries are a class of Galactic transients whose radio emission arises from their synchrotron-emitting jets.  The morphology and power of these jets correlates well with the X-ray spectral and variability properties of the source as it progresses through an outburst \citep{fender2004}.  
While some of the earliest radio studies of X-ray binaries showed that they can produce low-frequency radio emission in outburst \citep[e.g.][]{anderson1972, bash1973}, this class of sources has been relatively poorly characterised below 1\,GHz.  

The steady, compact jets observed at the beginning and end of an X-ray binary outburst are known to have flat radio spectra at GHz frequencies, with flux densities of up to a few tens of mJy \citep{fender2000}.  Theoretical models predict a low-frequency cutoff in the spectrum, either due to the internal shocks that accelerate the electrons having dissipated all of their energy \citep{malzac2014}, or to a change in the shape of the electron energy distribution from a power law to a Maxwellian \citep{peer2009}. 
By characterising the low-frequency compact jet spectrum through spectral fitting with simultaneous multi-wavelength observations, it is possible to test these models and use them to derive the physical properties of the jets. 
However, owing to the lack of sufficiently high-sensitivity low-frequency instruments, the low-frequency spectrum of X-ray binaries is still poorly constrained so the spectral cutoff has not to date been detected.

Closer to the peak of the outburst, X-ray binaries generate brighter, transient, relativistically-moving jets, which provide the best prospects for detection \citep[e.g.,][]{broderick2015,chandra2017}.  It is well established that standard expanding synchrotron bubble models \citep[e.g.,][]{vanderlaan1966} cannot explain the observed spectra and evolution of these transient jets \citep[e.g.,][]{broderick2018}.  However, to date there have been no sufficiently high-cadence, high-sensitivity studies done near outburst at low radio frequencies to test other models, which include slowed expansion \citep{hjellming1988} or continuous particle injection \citep[e.g.,][]{marti1992}.  
With well-sampled broadband radio spectra down to low frequencies, the above models can be tested, and 
physical parameters such as jet magnetic field strength can be determined from the location of the synchrotron self-absorption turnover \citep[e.g.,][]{miller-jones2004,chauhan19}.

As with any synchrotron transients, the low-frequency emission peaks later and at lower amplitude than higher-frequency emission, and the sensitivities of most instruments only enabled the detection of the brightest outbursts. However, the sensitivity of the Phase~II MWA coupled with its southern hemisphere location (accessing the Galactic Bulge, where a significant fraction of X-ray binaries are located) enables a more detailed exploration of the low-frequency properties of X-ray binary jets.

On larger scales, the jets interact with the surrounding interstellar medium, inflating cavities or lobes that can be filled with old, synchrotron-emitting electrons \citep{heinz2002}.
The excellent surface brightness sensitivity of the MWA Phase~II could help detect such lobes.  Furthermore, fits to the broadband spectra could help distinguish between free-free absorption and synchrotron self-absorption.  In the former case, this would provide information on the properties of the surrounding environment.
\newcommand{\dmu}{${\rm pc \ cm ^ {-3}}$\,}
\subsection{Pulsars}
%

Pulsar research at the MWA is facilitated by the  VCS \citep{2015PASA...32....5T} that records 100-$\mu$s/10-kHz resolution voltage data from 128 tiles, in dual-polarisation. These data can then be processed offline depending on the science application; for instance, they can be converted into power and summed to provide a modest sensitivity over a large FoV (i.e. incoherent beam), or coherently summed to make a phased-array (tied-array) beam on the target of interest for applications requiring high sensitivity. Notwithstanding the large data rates (28 ${\rm TB\,hr^{-1}}$) and processing requirements, the VCS has been exploited for a variety of science, including a low-frequency census of known (catalogued) pulsars \citep{xue2017} and 
high-resolution studies of pulsar scintillation and scattering 
\change{\citep[e.g.][]{2016ApJ...818...86B,2018ApJS..238....1B, Kirsten:2019}} and emission properties \citep[e.g.][]{McSweeney:2017,2017ApJ...851...20M}.
However, the data-transport constraints of VCS greatly limit monitoring and survey type  science, despite the large FoV. For instance, an all-sky survey for pulsars was formidable with Phase I due to the prohibitive cost of beamforming in order to attain high sensitivity across the full FoV. The advent of Phase~II with a compact configuration alleviates this major hurdle by virtue of the much larger synthesised beam. Further, the newly-developed capabilities such as the BL back-end and the  buffer mode of the VCS (Section~\ref{sec:digital_backends}), combined with the ability to reconstruct very high time resolution ($\sim$1$\mu$s) time series data open up prospects for undertaking detailed studies for uncovering the elusive pulsar emission mechanism. 

\subsubsection{All-sky pulsar surveys}

In the compact configuration, with the two hexes and the core providing a collecting area equivalent to that of the full Phase~I array, the processing cost for beamforming across the FoV is considerably reduced.
Specifically, the number of tied-array beams required to fill the FoV (at a gain level down to half power point) is reduced from $2.7 \times 10^5$ (Phase~I) to $3.9 \times 10^3$ (Phase~II compact). This corresponds to a reduction of more than two orders of magnitude in the computational cost, thereby making large-scale, high-sensitivity pulsar searches more tractable (and affordable) with the MWA. 

Large pulsar surveys have established histories of delivering high-impact science returns in the long run. A noteworthy example is the Parkes multibeam survey \citep{2001MNRAS.328...17M} that discovered the  double pulsar system J0737-3039A/B and the eccentric neutron star-white dwarf binary J1141-6545, both of which are proven laboratories for testing general relativity and alternate theories of gravity \citep{2006Sci...314...97K,2008PhRvD..77l4017B}. Its successor survey \citep{2010MNRAS.409..619K} found $>$200 pulsars, besides the very first population of FRBs \citep{2013Sci...341...53T}. As such,  conducting a full Galactic census of pulsars is a key science driver for the SKA and its pathfinders \citep{2015aska.confE..40K}. 
While the $\sim 1 - 2$\,GHz band proved efficient for probing deeper into the Galactic plane, low frequencies offer the advantages of higher survey speeds, and benefit from the increased brightness of pulsars.

\begin{figure}[t]
{\includegraphics[width=\columnwidth,angle=0]{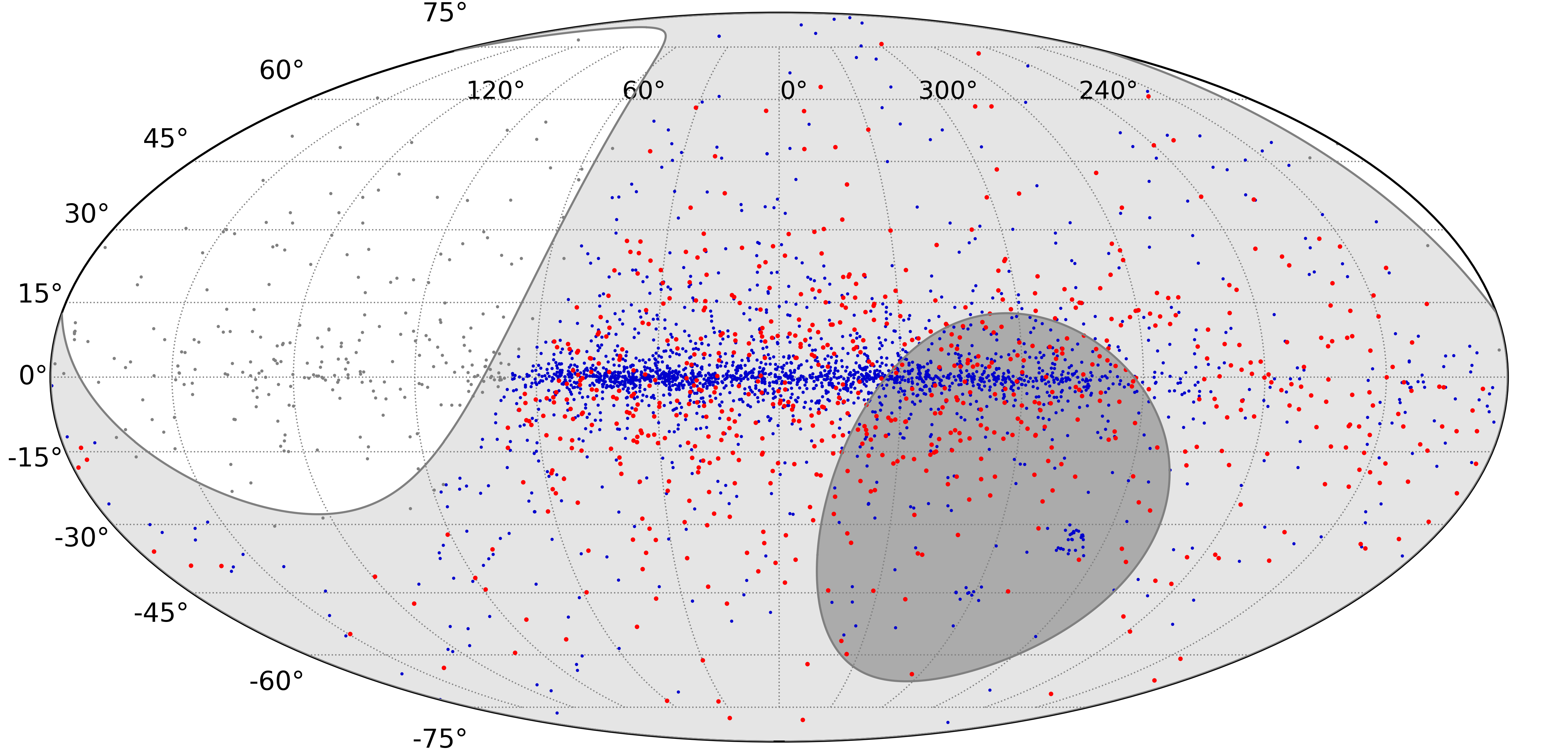}
}
\caption{
Simulated pulsar detections (red dots) from an all-sky high time resolution survey with the MWA. The grey region represents the MWA's visible sky, and the dark-grey region the sky  that is exclusively accessible by the MWA at frequencies below 300 MHz (declination $<-55^{\circ}$). The blue dots are pulsars from the ATNF pulsar catalogue v1.59 
} 
\label{fig:simpsrs}
\end{figure}


The above developments have led to the conception of the {\bf S}outhern-sky {\bf M}W{\bf A} {\bf R}apid {\bf T}wo-metre (SMART) survey -- an all-sky survey for pulsars with the Phase~II compact MWA, where the goal is to systematically survey the entire sky visible to the MWA (i.e. declination south of +30$^{\circ}$) at a frequency band of $140 - 170$\,MHz. 
A dwell time of $\sim 1.5$\,hr per pointing (maximum duration of uninterrupted recording feasible with the VCS) translates to a 10-$\sigma$ detection limit of $S_{150}$ $\sim 3$\,mJy (the mean flux density at a frequency of 150\,MHz), assuming a 30.72\,MHz bandwidth. This is more than a factor of two improvement\change{\footnote{\change{Assuming a spectral index of $-1.4$ and factoring in the loss of sensitivity at large zenith angles.}}} over the Parkes southern pulsar survey from 1990s (at a frequency of 430 MHz) which found 101 pulsars including 17 millisecond pulsars (MSPs) \change{\citep{parkes_sps, Manchester:1996}}. Searching at low frequencies will necessitate a large number of trial DMs, and will be limited by a ``pulse broadening horizon'', i.e. the distance beyond which temporal smearing from multipath propagation ($\tau _d$) limit (or degrade) the pulsar detectability. At 150\,MHz, this is $\sim 5$\,kpc ($\tau _d \sim 100$\,ms) for directions away from the Galactic disc \citep[cf.][]{2004ApJ...605..759B}. Pulsars with periods $\gtrsim 100$\,ms, DMs $\lesssim 300$\,\dmu and $S_{150} \gtrsim 3$\,mJy will be detectable with the MWA survey.  

Accurate modelling and forecasting of the expected yield from such a survey is however difficult, due to a number of unknowns, such as a possible spectral turnover at low frequencies, and temporal broadening from multipath scattering. It is further complicated due to the loss of sensitivity from projection (at large zenith angles) and the resulting non-uniformity in sensitivity (e.g. when observations are made in the drift-scan modes). Even then, a conservative forecast based on simulation studies  performed using the PSRPOPPy software \citep{bates2014}, after calibrating with recent MWA pulsar census work \citep{xue2017}, predicts $\sim$230 new pulsar discoveries (see Fig.~\ref{fig:simpsrs}) including $\sim$20-40 MSPs. 
Aside from this scientific promise, this will produce a digital archive of the full MWA-visible sky, and will serve as an important demonstrator survey for the low-band SKA.



\subsubsection{High-time resolution studies of pulsar emission}

\begin{figure}[t]
{
\includegraphics[width=.95\columnwidth, angle=-90]{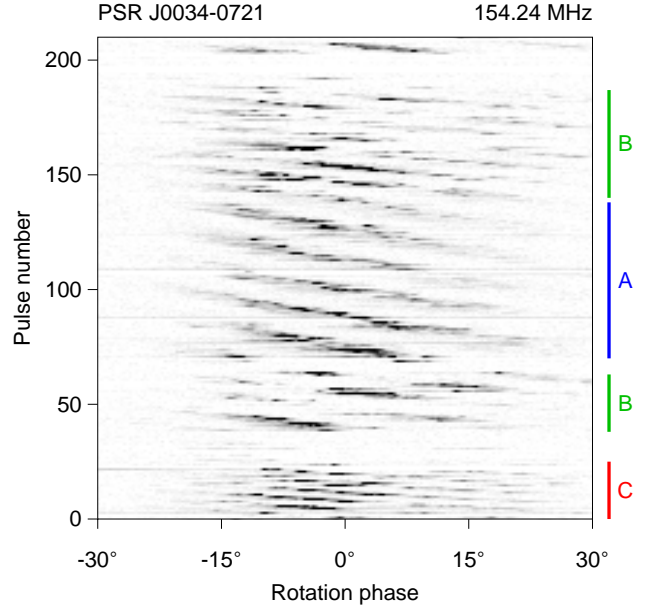}
}
\vskip -0.0cm
\caption{
A sequence of 210 individual pulses from PSR J0034$-$0721 that shows the phenomenon of sub-pulse drifting. The observations were made with Phase~II over the 140-170 MHz band, and processed using the newly-enhanced tied-array beam-former pipeline that resynthesises the VCS-recorded data to produce high-time resolution ($\sim$1\,$\mu$s) time series. The pulsar switches between three distinct drift bands (designated as Modes A, B and C) and exhibits the phenomenon of nulling (i.e. cessation of pulses) and is a promising target for studying emission mechanisms. 
} 
\label{fig:pulsestack}
\end{figure}

Pulsar emission phenomena occur on time scales spanning several orders of magnitude, with temporal structures lasting mere nanoseconds \citep[e.g.][]{Hankins:2003} to microseconds \change{\citep[e.g.][]{Hankins:1971,Cordes:1981, De:2016}} and even longer \citep[e.g.][]{Deshpande:2001, McSweeney:2017}. In general, smaller temporal structures map directly to smaller physical structures in the pulsar magnetospheres. High-time resolution studies of individual pulses from pulsars that exhibit both regular pulse-to-pulse variations such as sub-pulse drifting and microstructure (i.e. the finer components that make up individual subpulses) can thus provide important insights into the relationship between physical magnetospheric structures of different length scales. Such studies (which are still in their infancy) are now ripe to be explored with the new capabilities of the MWA-VCS's pulsar processing pipeline, where the beamformer pipeline \citep{Ord:2019} has been enhanced with an implementation that allows re-synthesising high-time resolution voltage time series by undoing the polyphase filter bank operation prior to the correlation stage. This allows attaining a time resolution down to $\sim$1\,$\mu$s, and is highly promising for studying short time-scale pulsar emission features at the low frequencies of the MWA (Fig.~\ref{fig:pulsestack}) where pulsars generally tend to be brighter. 

This new implementation has been successfully tested using observations of MSPs (Kaur et al. 2019). The ability to perform coherent de-dispersion on high-time resolution voltage time series allows the complete removal of the deleterious effects of temporal smearing caused by interstellar dispersion \citep{1975mcpr...14...55H}. Together with the large frequency lever arm provided by the MWA, this allows the measurement of pulsar dispersion measures with unprecedented levels of precision. The high quality pulse profiles that capture fine temporal structures by virtue of being able to perform phase coherent de-dispersion also significantly improves the precision with which the pulse arrival times can be measured. These new high-time resolution capabilities have wide-ranging applications from studies of millisecond pulsars to the exploration of a new parameter space in the quest to understand the pulsar emission mechanism.

\subsubsection{Sporadic and intermittent emission}

 While the emission from most pulsars is seen as regular sequences of pulses \citep[e.g.][]{McSweeney:2017}, some pulsars exhibit irregular emission, in the form of sporadic emission (e.g. giant pulses), or switching emission states \citep[e.g.][]{2006Sci...312..549K,2012ApJ...758..141L,2014MNRAS.442.2519Y,2017ApJ...851...20M,2018ApJ...869..134M,2019PASA...36...34M}. Such emission occurs on time scales of seconds to months, and the pulsars which exhibit such types of emission pose a major challenge to our understanding of the underlying physics that govern the radio emission mechanism. While there are several models that attempt to describe such intermittency \citep[e.g.][]{1985ApJ...299..917C,2008ApJ...682.1152C,2012MNRAS.419.1682J,2012ApJ...746L..24L,2013MNRAS.428..983S,2014MNRAS.437..262M}, none satisfy the enormous diversity of emission phenomenology observed. Furthermore, the sporadic and intermittent emission behaviour appears to be extremely broadband, based on  simultaneous  high-energy and wideband radio observations \citep[e.g.][]{2010ApJ...708.1254A,2013Sci...339..436H,2015ApJ...809L..12K,2017ApJ...851...20M,2018MNRAS.480.3655H}. In general, low-frequency coverage is lacking  for the vast majority of southern hemisphere pulsars, and in particular for  pulsars with sporadic or intermittent emission. Observations at low frequencies with the MWA are thus potentially promising to reveal emission characteristics that are substantially different to those observed at higher frequencies.

However, these sporadically emitting pulsars are inherently difficult to observe without regular, unusually long dwell times, which is generally difficult, and in particular not currently feasible using the VCS-like functionality.  The newly-developed voltage buffer mode can be suitably exploited to address this short-coming. 
This mode allows the MWA to circumvent the inherent limitation in studying intermittently emitting pulsars to some extent; for example, through suitable coordination with the observing programs that regularly observe such intermittent targets, such as UTMOST \citep{2017PASA...34...45B} and the Parkes radio telescope \citep[e.g.][]{2014MNRAS.445..320K}. The recent detection of low-frequency emission from the intermittent pulsar J1107$-$5907 \citep{2018ApJ...869..134M} provides an excellent demonstration of the MWA's unique advantage for conducting such coordinated broadband observations. As it is, broadband simultaneous observations involving multiple telescopes ostensibly provide valuable insights into the pulsar emission mechanism \change{\citep[e.g.][]{Bhat:2007, 2013Sci...339..436H, McSweeney:2019}}.  With the prospects of the VCS buffer mode and BL back-end becoming available for routine observations in the coming years, the Phase~II MWA will be promising for studying these sub-populations of pulsars.


\section{SOLAR, HELIOSPHERIC, AND IONOSPHERIC SCIENCE}
\label{sec:shi}
\subsection{Solar Imaging}
\begin{figure}
    \centering
    \includegraphics[width=\columnwidth]{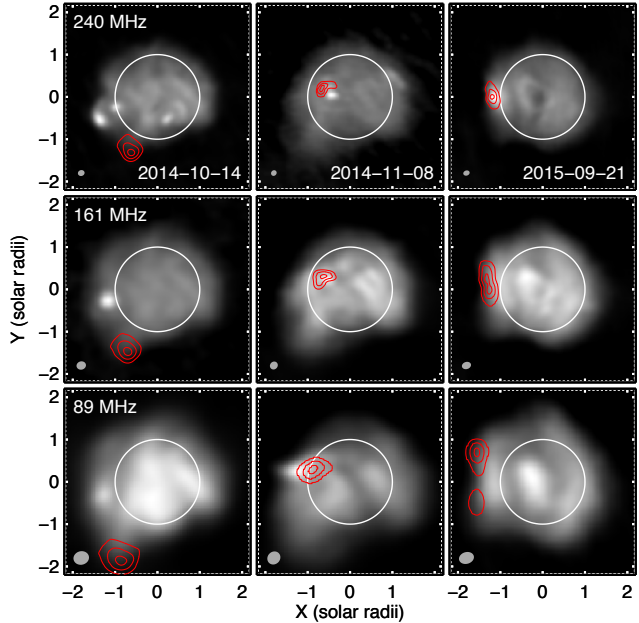}
    \caption{Examples of Phase~I solar observations at three frequencies on three different days. The grayscale images reflect the quiescent background over a 5-min period, and the red contours show type III bursts. Contour levels are at 30\%, 60\%, and 90\% of the peak intensity, and the total irradiance during burst periods exceeds the background by at least an order of magnitude. The solid white circle denotes the optical disk, and the gray ellipses represent the synthesised beam sizes. The burst site on 2015-09-21 (right column) elongates and ultimately splits into two components with decreasing frequency due to a diverging magnetic field structure \citep{McCauley17}. Structure can also be seen in the 2014-11-08 burst (middle panel), but it is difficult to interpret given the low spatial resolution.
    }
    \label{fig:shi}
\end{figure}

The defining characteristic of the MWA for solar observations is its outstanding snapshot monochromatic $uv$ sampling which leads to unprecedented imaging \change{dynamic range and} fidelity. 
\change{
To place this in perspective, historically, the imaging dynamic range provided by the state-of-the-art instrumentation has ranged between many tens to a few hundreds at best.
Some of the new generation interferometers being used for solar science, most notably LOFAR \citep{Mann:2011}, offer much higher angular and temporal resolution, and also operate at frequencies significantly lower than the MWA. 
Though the imaging quality remains limited due to the sparse $uv$ sampling, these data have enabled exploration of novel phase space and have lead to a variety of studies - ranging from those of active emissions like the type-III solar bursts \citep[e.g.][]{Morosan:2014} to the quiet corona \citep{Vocks:2018}, and studies underscoring the importance of coronal propagation effects \citep[e.g.][]{Kontar:2017, Chrysaphi:2018}.
In terms of their dynamic range, and the ability of pick up weak emission features, the MWA solar images now represent the state-of-the-art by a wide margin, as discussed in the following section.
}
Also important for capturing short-lived transient solar emission which can change rapidly in frequency and space, is the ability to select any 24 ``coarse'' 1.28\,MHz channels from the 80 -- 300\,MHz observing band.
These capabilities have been used to study not only the well-known strong radio bursts, but also much weaker nonthermal emissions, as well as more persistent features in the corona.

Figure~\ref{fig:shi} illustrates typical Phase~I MWA images of the Sun, both in a quiescent state and during a type III burst.
In both cases there are features that remain unresolved, particularly at the lower frequency. 
Thus the extended Phase~II MWA with a beam area reduced by a factor of 4 promises to shed new light on these phenomena. 

Below we summarise the work done with the Phase~I MWA, and outline how this work will be extended with the Phase~II MWA. These efforts are also leading toward the realization of SKA1-Low and its solar science goals \citep{nindos:2019}

\subsubsection{Calibration and imaging challenges.}
\label{sec:shi_cal_hdr}
At MWA frequencies, even the quiet Sun is orders of magnitude brighter than any Southern-hemisphere compact source, with radio bursts reaching flux densities many orders of magnitude brighter still.
This means that there is sufficient signal to noise to image the Sun easily, even when using a very short snapshot with extremely narrow bandwidth. 
However it also means that astrophysical sources are not typically visible when viewing the Sun unless extremely high dynamic range can be attained.
Consequently, a technique was developed for robust absolute flux density calibration using a sky model of diffuse Galactic emission \citep{Oberoi17}.
This technique exploits the MWA's wide field of view and numerous short baselines to which the Sun appears as an unresolved source.
Only a few baselines are required (and no imaging is necessary) making the technique computationally lean.
A prescription to transfer this flux calibration to solar images has also been developed \citep{Mohan17}. 

In order to follow the spatial, spectral, and temporal evolution of solar emission it is desirable to image the Sun with the maximum time and frequency resolution possible (typically 0.5\,s and 40\,kHz time and frequency resolution respectively).
However this can mean $\sim1\times10^5$\,images per minute of observation.
Therefore, in order to be able to analyse significant amounts of MWA data, it is imperative to use automated pipelines for calibration and imaging.
This has motivated the development of an \textit{Automated Imaging Routine for Compact Arrays for Radio Sun} \citep[AIRCARS; ][]{Mondal:2019}.
This end-to-end interferometric calibration and imaging pipeline also corrects for bandwidth decorrelation due to differing cable lengths, tunes the analysis to the needs of solar imaging, and is able to improve imaging quality well beyond what had been achieved by other fully-automated approaches.
AIRCARS has successfully been used in a completely hands-off manner for imaging data spanning a large range of solar conditions (peak brightness temperatures ranging from $10^6 - 10^9$\,K) and can routinely produce images with dynamic ranges between $10^3$ and $10^5$.

\subsubsection{Radio Bursts}
Type III bursts are caused by semi-relativistic electron beams that stream through and perturb the background plasma, generating Langmuir waves that ultimately produce intense radio emission.
The electrons are accelerated by magnetic reconnection during solar flares, which was directly evidenced by observations from the 32-tile MWA prototype \citep{Cairns18}, and the beams then propagate away from the Sun along magnetic field lines connected to the flare site.
In particular, this analysis showed definitive evidence for reconnection occurring in association with type III bursts and with X-ray emissions. 
Additionally, since the associations are not 1:1 in either direction, special conditions are required to produce observable radio emissions and X-rays. 

\citet{McCauley17} reported on Phase~I MWA observations of type III source regions that repetitively split apart in a direction roughly perpendicular to the inferred electron beam trajectory, and they concluded that this behaviour resulted from a distinctive magnetic field configuration implied by contemporaneous extreme-ultraviolet observations.
These results highlight the MWA's capability to probe electron beam trajectories, and thereby local magnetic field structures, in new and unexpected ways.
Type III bursts can also be used to probe the coronal density, which was explored by \citet{McCauley18}, who found much larger densities than expected from standard models.
This replicated decades-old findings from previous instruments, and MWA observations of the quiescent corona were used to estimate the extent to which the type III source heights were altered by radio propagation effects, which were found to be important but not entirely sufficient to explain the apparent density enhancements. 
Comparisons between observations of the scattered source (which includes any angular broadening and ducting), and simulations which typically ignore these effects offer one way to determine the impact of these processes.
Since some observations \citep{mercier_etal_2006} and theory suggest that sometimes scattering is very weak and sometimes strong \citep{subramanian_cairns_2011}, and the amount of scattering depends on a line integral involving the spectrum of density irregularities, these theory-data comparisons might remotely constrain the spectrum and spatial variations of density turbulence in the corona.

\citet{Mohan18} have succeeded in unambiguously disentangling the effects of scattering from intrinsic variations in emission morphology of a weak type III burst. 
This has \change{led} to the discovery of second-scale quasi-periodic oscillations in the angular size and orientation of the burst source with simultaneous oscillations in the intensity of the burst.
These observations cannot be explained in terms of the conventional Alfv\`{e}nic oscillations.
In the same data set \citet{Mohan:2019} also find an unrelated weak flare seen in the EUV and soft X-rays. 
Their analysis shows that this flare was responsible for some local coronal heating. A detailed study of the radio emissions associated with the magnetic loop undergoing heating has revealed a strong tendency for the observed short-lived narrow-band nonthermal radio emissions to cluster on a time scale of \SI{30}{\second}.
This clustering is very well formed and prominent when the heating event is in progress, but is seen to exist even prior to the start of the heating phase, suggesting that this time scale is independent of the heating event itself. 
This discovery was made possible by the high dynamic range spectroscopic snapshot imaging capability of the MWA, which enabled a reliable  detection of the weak nonthermal emissions originating from the source of interest, even in presence of other stronger activity. 
The improved point-spread-function of the MWA Phase~II, and its higher angular resolution will enable detection of even finer features and further increase the imaging dynamic range of MWA solar images.

Type II bursts are interpreted in terms of shocks accelerating electrons that generate Langmuir waves and then radiation at the electron plasma frequency and its harmonic \citep{nelson_melrose_1985,cairns_2011}.
Observational analyses and simulations predict that sometimes multiple regions of the shock produce the radio emission simultaneously or sequentially, depending on the event \citep{schmidt_cairns_2012b,schmidt_etal_2014, kozarev_etal_2015}.
MWA Phase~I observations of the 2014 September 7 type II \change{burst} show a stably-located, weakly/un-resolved source despite predictions for multiple type II subsources a factor of $\sim 4 - 10$ smaller than the beam size at $140$\,MHz.
The increased resolution of MWA Phase~II will provide a stronger test of the multiple-source hypothesis.  

Another window on these processes is provided by polarimetry. 
Studies of the polarized time- and spatially-varying source regions of type II and III bursts will also produce new science and constraints on emission mechanisms and scattering processes.

One long-standing issue is that the standard theory for type I, II, and III bursts involves generation of radiation near the electron plasma frequency that is $100\%$ circularly polarised in the sense of the ``ordinary'' (or ``o'') mode, yet this is not observed for all three: specifically, type Is are almost invariably $100\%$ circularly polarised, type IIs (except for herringbone fine structures) less than $\sim 20\%$, and type IIIs are occasionally up to $60\%$ but typically $\lesssim 20\%$ \citep{nelson_melrose_1985,suzuki_dulk_1985}. 
One interpretation is that the standard theories are wrong but another is that scattering depolarises the radiation \citep{melrose_2006}.
MWA Phase~II can address these issues by explicitly measuring the time- and spatially-varying sources in all $4$ Stokes parameters, searching for evidence of strongly polarised substructures that vary with time and space (e.g., depolarisation fronts moving away from the cores of subsources versus random flickering polarisation regions). 

In conclusion, studies of solar radio bursts are expected to benefit substantially from MWA-Phase~II's improvements in linear spatial resolution (a factor close to 2) and better $uv$ plane coverage (better imaging fidelity).
These should result in much better characterisation of the complex, sub-structured sources of type II and III bursts in position, time, and polarization, thereby enabling more quantitative tests of theoretical ideas and predictions. 

\subsubsection{Radio Counterparts of Coronal Mass Ejections}
The bulk of solar dynamics is dictated by solar magnetic fields, but beyond the photosphere their measurements have remained difficult. 
Radio observations form the only known techniques to measure these fields. The first successful attempt to model the spectrum of radio emission from the Coronal Mass Ejection (CME) plasma as gyrosynchrotron emission, and hence provide an estimate of the CME magnetic field was by \citet{Bastian:2001}. 
In spite of their well \change{recognised} merits, and considerable effort expended towards them, there are only a handful of successful detections available in the literature 
\citep{Maia:2007, Tun:2013, Bain:2014, Carley:2017}, some of which are different attempts to model the same CME. The key reason for this lack of success is that the intrinsic $T_B$ of gyrosynchrotron emission is at least a few orders of magnitude lower than that of the other nonthermal emissions usually present during the times when a CME has recently lifted off, and the imaging dynamic range provided by instrumentation available \change{until} recently fell short of the requirement to be able to image this weak emission. 
Recent work with the MWA has shown that with its improved imaging dynamic range, and ability to simultaneously sample its entire observing band, it should now be possible to routinely image the radio counterparts of CMEs, and model their spectra in a more robust manner than possible before \change{\citep{Mondal:2019b}}. 
The improved resolution of the MWA Phase~II will reveal the structures in CMEs in greater detail and its better imaging dynamic range will permit tracking of these emissions further out into the heliosphere, making these measurements relevant from a space weather perspective.

\subsubsection{Weak non-thermal emission}
Early observations with the 32-element MWA prototype revealed a previously unappreciated abundance of weak, short-lived (few s), narrowband (few MHz) impulsive nonthermal emission features carpeting the frequency--time plane \citep{Oberoi11}.
Detailed non-imaging investigations of weak impulsive nonthermal emissions have been carried out using the MWA Phase~I with the aim of providing robust statistics on these emissions.
Using an automated continuous wavelet transforms based approach, \citet{Suresh17} found that these impulsive emissions take place at the rate of many thousand per hour (as measured across 30.72\,MHz bandwidth).
Individual emission features were found to last for 1--2\,s and span 4--5\,MHz in bandwidth and $\sim 1 - 100$\,SFU\footnote{A solar flux unit or SFU equals 
$10^{4}$~Jy $= 10^{-22}$~W\,m$^{-2}$\,Hz$^{-1}$} in peak flux densities. 
Their characteristics suggest that these features might represent the weaker end of the distribution of solar type I bursts.
Using a Gaussian mixtures based statistical analysis technique, \citet{Sharma18} modeled the solar radio emission as a superposition of a slowly varying emission component of thermal origin and an impulsive, and hence nonthermal, component.
They estimated the flux density distribution as well as the prevalence of impulsive nonthermal emission in the frequency--time plane.
They found the fractional occupancy of nonthermal impulsive emission to lie in the range 17\%--45\% even during a period of medium solar activity, and that the flux density radiated in the impulsive component is very similar in strength to that radiated in the thermal emission-dominated slowly varying component. 

These studies were motivated by the desire to explore the use of weak nonthermal emissions in looking for signatures of weak magnetic reconnection events, the presence of which was originally suggested by \citet{Parker88}.
Referred to as the {\em nanoflare hypothesis}, the simultaneous presence of a large number of weak reconnection events carpeting the solar surface, but too weak to be detected individually, was proposed as a possible resolution of the coronal heating problem and remains one of the likely possibilities. 
At flux densities of $\sim 0.2$~SFU, these are the weakest nonthermal emissions reported at low radio frequencies using non-imaging techniques.
Ongoing imaging studies using MWA Phase~I suggest RMS variability during quiet sun conditions at the levels of $10^{-3}$~SFU (i.e. $10$~Jy), by far the weakest to be reported ever (Sharma et al., 2019, in preparation).
The high dynamic range techniques described in Sec.~\ref{sec:shi_cal_hdr} along with the higher angular resolution of MWA Phase~II will enable even deeper studies to come.
Though a lot remains to be done before their role in coronal heating can be elucidated, these studies have firmly established the presence and abundance of nonthermal emissions significantly weaker than were known before. 

\subsubsection{Quiescent Sun}
It is not only bursts that are of interest but also sources on the quiescent Sun, for instance associated with coronal holes, active regions, and the quiet Sun itself.
Work with MWA Phase~I intensity data shows new features of coronal holes and active regions \citep{McCauley17}. 
MWA Phase~II's better angular resolution and imaging fidelity will allow better characterisation of these sources and better comparisons with the SDO, SOHO, and RHESSI data from the UV to X-rays.

In addition, recent work using Phase~I data has produced the first low-frequency images of the quiescent solar corona in circular polarisation \citep{mccauley:2019}.
The Stokes $V$ structure at the lowest MWA frequencies is found to be generally well-correlated with the structure of the line-of-sight magnetic field component in a widely-used global magnetic field model at a height roughly corresponding to that of the radio limb.
Coronal magnetic field models are typically extrapolated from photospheric observations, and there are very few observations that can be used to constrain the field at the heights and scales probed by MWA observations.
This is a powerful and unique capability of the MWA that will be greatly enhanced by the improved spatial resolution of the Phase~II extended mode, as Phase~I could not resolve many of the features that would discriminate between competing models. 

\subsection{Observations of the Heliosphere}
\subsubsection{Interplanetary Scintillation}
\label{sec:IPS}
Interplanetary Scintillation (IPS) is the rapid ($\sim$1\,s) variability in brightness of compact ($\lesssim$1 arcsecond) sources due to scattering of the propagating wave by the solar wind: a supersonic outflow of turbulent plasma which fills interplanetary space.
A detection of IPS with the MWA was first made serendipitously in night-time astrophysical data by \citet{Kaplan15}.
Following this, a pilot study was made using observations designed for the purpose, at solar elongations most suitable for detecting IPS \citep{Morgan18}.
This pilot study demonstrated that the MWA has a unique capability for detecting IPS of several hundred sources simultaneously in a single 5-minute observation.
From a solar science perspective this means that the heliosphere can be mapped in unprecedented detail in a very short amount of time.

These findings motivated a 6-month long observing campaign, described in detail by \citet{Morgan:2019}.
The analysis of these data are ongoing, however much of the work so far has focussed on determining the astrophysical nature of the sources detected.
They differ somewhat from compact source populations at higher frequencies, most notably they are dominated by peaked-spectrum sources \citep{Chhetri18}.
The precise counts of the various subpopulations are described in \citet{Chhetri18b}.
\citet{Sadler:2019} have shown that the most compact IPS sources are high-redshift sources, 1/3 of which have a redshift $>2$.

A further key finding \citep{Morgan18} was the necessity to maximise the sensitivity and imaging fidelity on short timescales ($\sim$1\,s).
In contrast to most astrophysical applications, where MWA sensitivity is limited by confusion noise \citep[see e.g.][]{GLEAM}, IPS sources are more rare and therefore the confusion limit is at a considerably lower flux density.
However, since IPS detections are made by measuring an increased \emph{variance} in flux density, the detection limit in an observation only reduces with the \emph{fourth} root of observing time.
This makes maximising the instantaneous sensitivity of the array critical.

This can be maximised in synthesis images by choosing a natural weighting scheme (where all baselines are weighted equally).
However this comes with two disadvantages: firstly the resolution is poor compared to a weighting scheme which increases the weight of longer baselines (i.e. a uniform weighting scheme); secondly the PSF of the array in a naturally weighted image typically has higher sidelobes.
With the Phase~I MWA, this latter effect is very strong due to the very high concentration of tiles in the core. 
This is in contrast to uniformly-weighted images which have an exceptionally good PSF due to the large number of interferometer elements in the MWA and their pseudo-random arrangement.
For IPS observations (where the Sun is typically in the sidelobes), uniform weighting has the added advantage that it downweights the extended quiet Sun significantly.

As shown in \cite{Wayth:2018}, by utilising the extended Phase~II MWA and using natural weighting, we combine the resolution and exceptionally low sidelobes of uniformly-weighted Phase~I images with the sensitivity of naturally-weighted Phase~I images).
This should lead to a factor-of-two improvement in sensitivity \citep{GLEAM}.

\subsubsection{Faraday Rotation Measurements}
Kinetic and magnetic energy from a CME can create major, potentially damaging disturbances in the near-earth environment, and there is high interest in being able to predict both the arrival time and the effects of CME impacts with sufficient precision and lead-time.
Since the geo-effectiveness of CME impacts is strongly dependent on the orientation of the magnetic field in the CME, a long-standing goal of low-frequency observations in the context of space weather is to measure the orientation of the magnetic fields in propagating CMEs.  The only known means of remote detection of this orientation is via the measurement of Faraday rotation (FR) in the plasma of the CME \citep{Oberoi12}.

Because of the radial dependence of coronal/solar wind density and magnetic field strength, attempts to detect Faraday rotation from heliospheric plasma, and in particular from CME events, have been confined to regions close to the sun.  Some measurements have used linearly polarized signals from spacecraft \citep[e.g. the Helios spacecraft at distances of 2--20 solar radii, using the \SI{2.295}{\giga \hertz} carrier or the MESSENGER spacecraft using the 8\,GHz carrier, probing the corona at a few solar radii][]{holweg_1982,jensen_2013,efimov_2015}.  Others have used pulsars \citep{Ord_2007,You_2012}, again at small solar elongations where the coronal plasma and field strengths are high.  Observations of polarized extragalactic sources have also been conducted \citep[e.g.][]{lechat_2014}, most recently with considerable success using the upgraded Karl G. Jansky VLA \citep{Kooi_2017}, also at small solar elongations.  The pulsar and extragalactic source observations have typically been done at low GHz frequencies.

In order to probe heliospheric and CME Faraday rotation at large fractions of an au, much lower frequencies must be used so that the amount of rotation is detectable.  This requires that a source of background linearly polarized emission must be present, but discrete polarized sources tend to be few and weak at MWA frequencies.
Furthermore, to be of practical utility, any FR measurements must target the inner heliosphere for CMEs which may be on a collision course, which generally means performing the FR observations during the daytime.
Due to the strength of solar radio emission, this places extraordinary demands on the imaging dynamic range of these observations.

The MWA Phase~I system has allowed major advances in these studies for two key reasons.
First, as has been demonstrated by MWA observations \citep{Lenc2016}, the polarized galactic synchrotron background has much more power on large angular scales than on small scales, a discovery made possible by the excellent short-baseline $uv$ coverage and imaging capability of the MWA.
Thus, by employing low angular resolution imaging, a strong polarized signal against which to measure FR is readily available in essentially all directions.
This is a dramatically more favorable scenario than trying to use a sparse grid of weak polarized discrete sources.
Second, the excellent instantaneous $uv$ coverage of the MWA combined with thousands of independent closure quantities has allowed the development of very high dynamic range solar imaging, which in turn creates a realistic possibility of detecting much weaker emission (the polarized Galactic background) during the daytime, and making FR measurements of CME plasma in the inner heliosphere.
Work continues to further improve dynamic range via the development of direction-dependent ionospheric calibration on small linear scales.

For FR work, the Phase~II system differs from Phase~I in one important respect, namely that for either high or low angular resolution work, essentially all 128 tiles are available, instead of a subset.
For low-resolution FR studies, the compact configuration thus yields not only greater sensitivity, but also denser $uv$ coverage with extensive redundancy (by design) to constrain calibration solutions.
Improved calibration, the inclusion of direction-dependent, small linear scale ionospheric calibration, and the much simpler nature of the angular structure of the solar emission on scales of 0.5 degrees and larger, combine to offer the prospect of meaningful daytime FR measurements.
Such measurements are potentially possible using existing Phase~I data, but compact configuration Phase~II data will be inherently superior.

The first detection and characterization of spatially resolved FR associated with an interplanetary CME would be a major result, offering a densely sampled large-volume view of a quantity directly related to the magnetic field geometry.
This would transform studies of CME dynamics, and would point the way to a future network of low frequency imaging arrays, modeled after the MWA, that could form a crucial part of a space weather prediction network.

\subsection{Observations of the Ionosphere}
During Phase~I, ionospheric research has been driven largely by the needs of individual science teams to measure and compensate for the effects of the ionosphere on their own observations \citep[e.g.][]{Loi2015c,jordan17}. 
Additionally, \citet[][]{Arora2015,Arora2016} have investigated the possibility of using GPS measurements to assist in low-frequency radio astronomy calibration.

This work has shown that the refractive shifts of radio sources at MWA frequencies allows measurements of the density gradient of the ionosphere with unprecedented accuracy; and (thanks to the sensitivity and wide field of view of the MWA) with an unprecedented number of pierce points.
These practically-motivated investigations have nonetheless led to unexpected discoveries.
Most notably, \citet{Loi2015a} discovered strikingly regular linear features in the observed refractive shifts in radio sources.
Through careful investigations, including splitting the array to measure distance via parallax, it was shown that these features are located within the magnetosphere and follow the Earth's magnetic field lines.

More recently, ongoing work using AIRCARS under a diverse range of solar conditions has lead to the conclusion that the variations in the ionospheric phase across the $\sim$0.5$^{\circ}$ radio sun are the next most constraining bottleneck limiting the imaging dynamic range \citep{Mondal:2019}. 
The large flux density of the Sun is in such a high SNR regime that solar imaging is sensitive to small changes in the relative Total Electron Content (TEC): $\sim$1\,mTECU (where 1\,TECU = 10$^{16}$\,electrons\,m$^{-2}$) over small spatial ($\sim$few 100\,m) and temporal ($\sim$10\,s) scales.
In calibrating out these effects we will directly measure the ionosphere on much smaller scales than previous MWA work, but with the same unprecedented accuracy.

\section{OTHER OPPORTUNITIES}
\label{sec:other}
While the majority of MWA collaboration efforts fall under the four primary science themes, the instrument's flexibility offers other opportunities as well. Here we highlight new directions that have developed since the original conception of the telescope.

\subsection{Space Situational Awareness}
The MWA was first used to detect objects in low Earth orbit (LEO) while the array was still being commissioned \citep{2013AJ....146..103T}.
This initial investigation used correlated data to image the sky and detect FM radio broadcasts reflected from the international space station (ISS) as it transited over Western Australia.
In a study designed to detect radio emission from fireballs, \citet{2018MNRAS.477.5167Z} serendipitously detected numerous objects in LEO, including the ISS (100\,m diameter), the Advanced Land Observing Satellite (10\,m), the defunct satellite Alouette-2 (1\,m), and the cube-sat Duchifat-1 (10\,cm).
By mapping the location of LEO objects during their passage through the field of view of the MWA, it is possible to recreate their orbital parameters.
With the resolution of the Phase~I MWA, objects in LEO can be located to well within the nominal 1\,km error ellipse that accompanies the orbital parameters listed in the TLE catalogue.
The detection of objects smaller than the ISS validate the predictions of \citet{2013AJ....146..103T}, and demonstrate the capability of using the MWA to both monitor and detect objects in LEO, including space debris, allowing the MWA to contribute to Australian and global Space Situational Awareness (SSA).

The MWA has been used as the receiving station of a passive bi-static radar system, to again detect the ISS as well as aircraft and meteors trails \citep{palmer_Surveillance_2017}.
In the passive bi-static radar set-up, terrestrial broadcast stations (in the FM range $80-108$\,MHz) are used as illuminators of opportunity.
The MWA records baseband data using the VCS and forms beams on the expected location of the ISS, and in the direction of the Geraldton or Perth based stations.
This is the set-up used by \citet{palmer_Surveillance_2017}, however the maximum expected degree of correlation between the direct and reflected signals is less than unity due to the corrupting effects of the ionosphere as it diffracts the direct signal over the horizon and into the field of view of the MWA.
Using data captured at sites local to the FM broadcast stations in Perth has been shown to work well, and increase the degree of correlation between the direct and reflected signals \change{\citep{Hennessy:2019}}.
The use of radar techniques to detect and monitor objects in LEO means that the MWA can not only determine the location of the objects, but the line of sight distance, velocity, and acceleration.

LEO constitutes orbits of less then $2,000$\,km, whereas the longest baselines of MWA Phase~I have a near/far field transition that is at $4,800$\,km (at $80$\,MHz).
In the work of \cite{2018MNRAS.477.5167Z}, only baselines shorter than \SI{387}{\meter} were imaged to ensure satellites and meteors were in the far field of the MWA. Because not all Phase~I baselines were able to be used, the final sensitivity was reduced. 
The Phase~II compact configuration of the MWA contains many more short baselines than the Phase~I configuration, and will therefore translate to an increase in sensitivity for observing objects in LEO, and an increased capability for SSA.
\subsection{Search for Extraterrestrial Intelligence}
\label{sec:seti}

The search for extraterrestrial intelligence (SETI) has been reinvigorated in recent years, in part by the realization that many stars are orbited by planets with conditions suitable for life as we know it, and also by the launch of major new SETI initiatives such as Breakthrough Listen \citep[BL;][]{2017Wordenetal}, which is currently undertaking a pilot project at MWA (Croft et al.\ in prep).

The MWA has several characteristics that make it ideal for SETI surveys, including:
\begin{itemize}
\item A wide field of view, so that a typical pointing contains many targets of interest, and so that much of the sky is covered to a depth of at least several hours each year
\item Flexible digital backends, enabling commensal surveys (beamforming for SETI without the need to control the primary beam pointing) and the generation of high resolution spectra from low-level raw voltage data 
\item An extremely radio-quiet site, minimizing the major contaminant in any SETI survey --- human-generated radio frequency interference (RFI)
\item Access to an under-explored region of the radio spectrum that may be very promising for the detection of bright extraterrestrial transmitters similar to those produced by human civilization \citep{loeb07}.
\end{itemize}

The MWA has already undertaken SETI experiments \citep{tingay:center,tingay:anticenter,tingay:omm} using data cubes generated by the correlator that were gathered for other primary science goals, and subsequently searched for candidate SETI signals. Similar studies could be done in Phase~II (with improved angular resolution), but the new capabilities of the array can provide additional advantages. SETI is one of a handful of science areas that was not covered in the original MWA science paper \citep{Bowman:2013}, but has now become a substantial area of research at MWA in part as a result of experience gained during Phase~I.



The improved angular resolution in the MWA Phase~II's extended configuration will allow more precise localization of candidate SETI signals. Higher spectral resolution from the correlator improvements currently underway, along with real-time beamforming planned as part of the BL engagement, will enable better fidelity both for narrow-band signals that have been the target of the majority of previous SETI searches, as well as better classification of more complex and broad-band signals. In addition to allowing a more powerful SETI search, these improvements may enable better RFI detection, classification, and rejection. 
\subsection{Archive Analysis}
\label{sec:archive}

The MWA data archive provides public access to all raw visibility data and a subset of VCS data from both Phase~I and Phase~II of the instrument.  The archive is accessible through the All-Sky Virtual Observatory (ASVO) at \url{http://asvo.mwatelescope.org} using a web dashboard or a Python-based API or commandline client.  Data is typically made public on the archive 18~months after its initial collection.  The archive contains 12 Petabytes (PB) of Phase~I data spanning mid 2013 to mid 2016, including 2.5~PB of VCS data.  It currently contains 18~PB of Phase~II data collected so far since 2016 October 1, of which approximately 70\% is from the extended array configuration and 4~PB is VCS data.   The AVSO interface has completed approximately 45,000 job requests to date and served over 1.3~PB to users.

Archived Phase~II data was recently used by \citet{tingay:omm} to perform an after-the-fact search for emitted signals from 1I/2017 U1 `Oumuamua to test the speculative possibility that the object is associated with extraterrestrial intelligent life.  The study used serendipitous observations identified in the archive by comparing the orbital trajectory of `Oumuamua with the target field positions and acquisition times of archived data.  An 86-second dataset was found that overlapped with the `Oumuamua trajectory in two months of archived data spanning 2017 November 1 through 2018 January 10, when the search was performed.

Similarly, \citet{croft:2016} used archival MWA data to search for radio transient emission coincident with two candidate high-energy neutrino events detected by the ANTARES neutrino telescope. No counterparts were detected, but they were able to place upper limits on progenitors in host galaxies within the localization of the neutrino events (\SI{10e37}{\erg\per \second} at \SI{20}{\mega\parsec}) and redshift limits if the source was not from the nearby galaxies ($z\geq0.2$). 

The large field of view of the MWA greatly increases the likelihood that the data archive contains observations of a given part of the sky at any time.  The typical MWA field size of 500 to 1000\,deg$^2$ translates to approximately 2.5 to 5\% of the southern sky observed at any given time.  This makes the MWA archive more effective for follow up of transient events compared to data archives of dish-based telescopes with smaller fields of view. Further, the archive contains observations of the entire sky below +30 declination, as well as thousands of hours of repeated observations for a subset of the sky covering more than 2000 deg$^2$ that is revisited nightly for months.

\section{CONCLUSION}
\label{sec:conclusion}
The capabilities of the MWA afforded by compact and extended array configurations, new digital back-ends, and a rapid-response triggering system offer opportunities for improved and new science programs. We have outlined here many ways in which the four original science themes of the MWA collaboration will continue to explore the low frequency astrophysical sky with the Phase~II upgrade. In addition, we outlined new directions outside the original themes. As an SKA low-band precursor, the MWA has and continues to be a rich test bed for science and technical demonstrations.

The Phase~II upgrade has added to the MWA's strength as a flexible multi-purpose instrument. 
The extended configuration has greatly enhanced the potential for high resolution galaxy surveys and foreground modeling for the EoR studies, while the compact configuration yields very high sensitivity to diffuse emission and a unique hybrid layout to test analysis techniques. 
The new rapid-response triggering system makes the MWA an excellent instrument for followup observations of transient events triggered from EM and multi-messenger signals.

With the MWA's versatile design, flexible observing modes,  Open Skies policy, and public data access through the All-Sky Virtual Observatory, it will continue to serve the astronomical community and produce fruitful scientific results for years to come.

\begin{acknowledgements}
The MWA Phase~II upgrade project was supported by Australian Research Council LIEF grant LE160100031 and the Dunlap Institute for Astronomy and Astrophysics at the University of Toronto. 
Parts of this research were supported by the Australian Research Council Centre of Excellence for All Sky Astrophysics in 3 Dimensions (ASTRO 3D), through project number CE170100013.
APB is supported by an NSF Astronomy and Astrophysics Postdoctoral Fellowship under award AST-1701440. CMT is supported by an ARC Future Fellowship under grant \change{FT180100321}.
GEA is the recipient of an Australian Research Council Discovery Early Career Researcher Award (project number DE180100346).
SC and the pilot Breakthrough Listen SETI program are supported through the Breakthrough
Initiatives, sponsored by the Breakthrough Prize Foundation.  D.K. was additionally supported by by NSF grant AST-1816492.
GG acknowledges the postdoctoral research fellowship by CSIRO.
\change{SWD acknowledges an Australian Government Research Training Programme scholarship administered through Curtin University.}

This scientific work makes use of the Murchison Radio-astronomy Observatory, operated by CSIRO.
We acknowledge the Wajarri Yamatji people as the traditional owners of the Observatory site. 
Support for the operation of the MWA is provided by the Australian Government (NCRIS), under a contract to Curtin University administered by Astronomy Australia Limited.
We acknowledge the Pawsey Supercomputing Centre which is supported by the Western Australian and Australian Governments.
\end{acknowledgements}

\bibliographystyle{pasa-mnras}
\bibliography{MWA-II_science,time_domain,eor_bib,SHI_bib,seti,geg}
\end{document}